\documentstyle[12pt]{article}
\newcommand{\bce}{\begin{center}}
\newcommand{\ece}{\end{center}}
\newcommand{\beq}{\begin{equation}}
\newcommand{\eeq}{\end{equation}}
\newcommand{\bea}{\vspace{0.25cm}\begin{eqnarray}}
\newcommand{\eea}{\end{eqnarray}}

\newcommand{\ba}{\begin{array}}
\newcommand{\ea}{\end{array}}

\def\lsim{\mathrel{\rlap{\lower4pt\hbox{\hskip1pt$\sim$}}
    \raise1pt\hbox{$<$}}}	  
\def\gsim{\mathrel{\rlap{\lower4pt\hbox{\hskip1pt$\sim$}}
    \raise1pt\hbox{$>$}}}	  

\def\lsim{\mathrel{\rlap{\lower4pt\hbox{\hskip1pt$\sim$}}
    \raise1pt\hbox{$<$}}}         
\def\gsim{\mathrel{\rlap{\lower4pt\hbox{\hskip1pt$\sim$}}
    \raise1pt\hbox{$>$}}}         

  \advance\oddsidemargin  by -1in
  \advance\evensidemargin by -1in
  \advance\topmargin      by -0.5in
\def\lsim{\mathrel{\rlap{\lower4pt\hbox{\hskip1pt$\sim$}}
    \raise1pt\hbox{$<$}}}         
\def\gsim{\mathrel{\rlap{\lower4pt\hbox{\hskip1pt$\sim$}}
    \raise1pt\hbox{$>$}}}         

\def\beq{\begin{equation}}
\def\endeq{\end{equation}}
\def\arr{\begin{eqnarray}}
\def\endarr{\end{eqnarray}}
\makeindex

\begin{document}


\phantom{.}{\large \bf \hspace{9.4cm} KFA-IKP(Th)-1994-20\\
\phantom{.}\hspace{11.9cm}1 June 1994\\ }
\begin{center}
{\LARGE \bf  Color transparency \\
after the NE18 and E665 experiments: \\
Outllok and perspectives at CEBAF.$^{*}$ \vspace{0.5cm}\\}
{\large \bf J.Nemchik$^{1,3)}$,
N.N.Nikolaev$^{2,3}$
and B.G.Zakharov$^{2}$\bigskip\\ }
{\sl
$^{1)}$Institute of Experimental Physics, Slovak Academy of
Sciences, \\Watsonova 47, 043 53 Kosice, Slovak Republik   \\
$^{2)}$L.D.Landau Institute for Theoretical Physics, GSP-1, 117940,\\
ul.Kosygina 2, V-334, Moscow, Russia\\
$^{3)}$IKP(Theorie), KFA J\"ulich, D-52425 J\"ulich, Germany
\vspace{0.4cm}\\ }

{\bf ABSTRACT}
\setlength{\baselineskip}{2.6ex}
\end{center}
{CEBAF is a high-luminocity factory of virtual photons with
variable virtuality $Q^{2}$ and transverse size. This
makes CEBAF, in particular after the energy upgrade to (8-12)GeV,
an ideal facility for uncovering new phenomena, and opening
new windows, at the interface of the perturbative and
nonperturbative QCD. We discuss color transparency as the case
for a broad program on electroproduction of vector mesons
$\rho^{0},\,\omega^{0},\,\phi^{0}$ and their radial excitations
$\rho',\,\omega',\,\phi'$ at CEBAF. We also comment on the
second generation of experiments on color transparency in
$^{4}He(e,e'p)$ scattering, which are also feasible at CEBAF.

In 1994, we can make more reliable projections into future
because our
understanding of the onset of color transparency has greatly
been augmented by two experiments completed in 1993:\\
i) no effect of CT was seen in the SLAC NE18 experiment on
$A(e,e'p)$ scattering at virtualities of the exchanged photon
 $Q^{2} \lsim 7$ GeV$^{2}$, \\
ii) strong signal of CT was observed in the FNAL E665
experiment on exclusive $\rho^{0}$-meson production in deep
inelastic scattering in the same range of $Q^{2}$. \\

We discuss the impact of these observations on the CEBAF
experimental program. We argue they both are good news, both
were anticipated theoretically, and both rule in the correct
QCD mechanism of the onset of CT.
} \bigskip\\
\begin{center}
{\bf \large $^{*)}$~Presented by NNN at the Workshop on
CEBAF at Higher Energies\\
CEBAF, 14-16 April 1994}
\pagebreak\\
\end{center}


\section{Introduction}


The fundamental prediction of QCD is that the quark configurations
with small transverse size $\vec{r}$ have small interaction
cross section [1], which was dubbed {\sl color transparency}
(CT) [2]. Looking for CT is long discussed as the case for the
high-luminocity, high-duty cycle, (10-20) GeV electron facility,
which is well documented in the ELFE project [3] (ELFE $=$
European Laboratory for Electrons). In the meantime, the good
news from CEBAF is a possibility of the (8-12)GeV upgrade, which
opens exciting possibilities of doing CT physics at CEBAF.

CEBAF is a high-luminocity factory of virtual photons. Higher
energy means a higher virtuality $Q^{2}$ of photons, and higher
$Q^{2}$ means that smaller sizes are becoming accessible  reaching
eventually into
the perturbative QCD region. Higher energy also means
longer lifetime of these small-size
states. However, from the very start,
we must emphasize that as far as CT physics is concerned,
the {\it purely perturbative} region lies well beyond the
kinematical range of CEBAF experiments on {\it exclusive} processes,
even after the (8-12)GeV energy upgrade.
 Testing the {\it purely perturbative} QCD
to few decimal places is a task of {\sl inclusive} experiments at
superhigh energy
facilities like LEP or HERA. Even at LEP and HERA, the predictive
power of the {\it purely perturbative} QCD rapidly deteriorates
when the {\it exclusive} processes are considered. The real task
of the CEBAF experiments is {\it to uncover new QCD
phenomena in exclusive reactions at the interface of the perturbative
and nonperturbative QCD}. Very conservative conclusion of this
overview is that the energy-upgraded CEBAF shall do the job.

Before jumping into conclusions on the feasibility of CT physics
at CEBAF, one must recall and critically summarize the results
of the two CT experiments completed in 1993:
\begin{itemize}
\item
The $A(e,e'p)$ reaction on the $D$, $C$, $Fe$ and $Au$ targets
was studied by the SLAC NE18 collaboration with the negative
result: no CT effects are seen at $Q^{2} \leq 7$ GeV$^{2}$ [4].

\item
The FNAL E665 experiment [5] on exclusive production of
the $\rho^{0}$ mesons in deep inelastic scattering of muons
on nuclei produced a solid evidence for CT in precisely the same
range of $Q^{2}$ as explored in the NE18 experiment.

\end{itemize}

The early history of CT focused on the quasielastic $A(e,e'p)$
scattering of electrons on nuclei. A number of predictions of
precocious CT at low $Q^{2}$ were published (for the review
and references see [6]), and the
failure to confirm this precocious CT in the NE18 experiment
considerably dampened the whole subject of CT. Fortunately,
more consistent treatment rather {\it predicted} a very slow
onset of CT in the $A(e,e'p)$ scattering [7,8]. As a matter of
fact, the NE18 results do perfectly confirm the correct theory
and rule in the mechanism of CT, which is alive and well, and
we can joyfully recite Mark Twain's telegram to the Associated
Press: {\it \bf "The reports of my death were an exaggeration".}

The parallel development was a theory of CT in (virtual)
photoproduction of vector mesons $\gamma^{*}N\rightarrow VN$.
{}From the theoretical point of view, this is a much cleaner
case, with a well understood shrinkage of the transverse size
of the virtual photon with the increase of $Q^{2}$ [9,10].
The prediction [9], not a postdiction,
of the precocious onset of CT was confirmed
by the FNAL E665 experiment [5], which put the CT physics in the
right perspective.

The strong point which we wish to make in this review is that
after the energy upgrade, CEBAF experiments on exclusive
electroproduction of vector mesons can significantly contribute
to our understanding of the onset of CT. Furthermore, the
experiments on production of the radially excited vector
mesons will open an entirely new window not only
on the mechanism of CT,
but also on the quark structure and poorly known spectroscopy
of radial excitations. In the $A(e,e'p)$ sector, we comment on
the potential of experiments on the $^{4}He$ target, in which
the onset of CT is sooner than for any
other nucleus owing to the small
size of the $^{4}He$ nucleus.

In this contribution to the Workshop on CEBAF at Higher Energies
we concentrate on the recent experimental and theoretical
developments, for the earlier reviews on the subject see [3,11-14].


\section{CT in exclusive production of vector mesons}


\subsection{CT and dipole cross section}

In order to be quantitative, let us set up the theoretical
framework, which is the lightcone dipole-cross section
representation [15]. Mesons can be viewed as color dipoles.
The distribution of the transverse size $\vec{r}$ of color
dipoles in the meson is given be the $q\bar{q}$ wave function
$\Psi(z,\vec{r})$, where $z$ is the fraction of meson's
momentum carried by the quark. This mixed $(z,\vec{r})$ lightcone
representation is custom tailored for description of CT.
By the Lorentz-dilation, in the high-energy scattering the dipole
size  $\vec{r}$ becomes as good a conserved
quantum number as an angular momentum.
The fundamental quantity which describes all the scattering
processes, is the dipole
cross section $\sigma(\nu,r)$ for interaction
of the color dipole of size $r$ with the target nucleon.
Of course, apart
from the $q\bar{q}$ Fock state, the lightcone hadrons contain
the higher $q\bar{q}g....$ Fock states. The effect of gluons in
the {\it projectile} dipole brings in the dependence of
$\sigma(\nu,r)$ on energy $\nu$, which can be related to the gluon
structure function $G(x,q^{2})$ of the {\it target nucleon} [16,17].
Specifically, at small $r$, the dipole cross section is
$\propto r^{2}$ ,
\beq
\sigma(\nu,r)={\pi^{2} \over 3}r^{2}\alpha_{S}(r)G(x,q^{2})
\label{eq:2.1.1}
\endeq
where $q^{2}= A_{0}/r^{2}$ with $A_{0}\approx 10$ [18], and
$x=q^{2}/2m_{p}\nu$. This CT property $\propto r^{2}$ derives
from the decoupling of gluons from a very small color dipole.
At large $r\gsim 1$f, the dipole cross section saturates
because color forces do not propagate beyond the confinement
radius. At high energy, the dipole cross section is a solution
of the generalized BFKL equation [19] and describes, for instance,
structure functions of deep inelastic scattering at HERA [20].
For any $q\bar{q}$ state, the total cross section of interaction
with the target nucleon equals
\beq
\sigma_{tot}
=\int_{0}^{1} dz \int d^{2}\vec{r}\,\sigma(\nu,r)
|\Psi(r,z)|^{2} \,.
\label{eq:2.1.2}
\endeq
The dipole cross section is a universal function of $r$, the
dependence on the
process is only contained in the wave function of the $q\bar{q}$
state. Fig.~1 shows qualitatively, how the dipole cross section
is probed in different processes:
\begin{itemize}
\item
The total pion-nucleon cross section $\sigma_{tot}(\pi N)$ and
the real photoabsorption cross section probe the region $r\gsim 1$f.
\item
The $\sigma_{tot}(J/\Psi N) \approx $(5-6)mb comes from $r \sim
R_{J/\Psi} \approx 0.45$f.
\item
The real photoproduction of the open charm probes the region of
$r\sim 1/m_{c} \approx 0.15$f.
\item
The proton structure function $F_{2}(x,Q^{2})$ receives contribution
from $1/\sqrt{Q^{2}} \lsim r \lsim 1$f, and CT property of the
dipole cross section is crucial for the Bjorken scaling [12,15,17].
\item
The scaling violations in $F_{2}(x,Q^{2})$ are dominated by the
contribution from $r\sim 1/\sqrt{Q^{2}}$.
\end{itemize}
In CT experiments one looks for a weak intranuclear final (and initial)
state interactions, which will be the case if the nuclear production
amplitude is dominated by the dipole cross section at small $r$
such that $\sigma(\nu,r)$ is much smaller than the free-nucleon
cross section. Whether the particular exclusive
reaction is selective of such a small
$r$ or not, requires the case-by-case study.



\subsection{How CT is tested in leptoproduction of
vector mesons?}

The exclusive (elastic) production $\gamma^{*}p\rightarrow Vp$
($V=\rho^{0},\phi^{0}, J/\Psi$,...) is an ideal laboratory for
testing CT ideas [9,10,12,21-24].
The forward production amplitude equals
\beq
M=\langle V|\sigma(\nu,r)|\gamma^{*}\rangle
=\int_{0}^{1} dz \int d^{2}\vec{r}\,\sigma(\nu,r)\Psi_{V}^{*}(r,z)
\Psi_{\gamma^{*}}(r,z)
\label{eq:2.2.1}
\endeq
Here $\Psi_{\gamma^{*}}(r,z)$ is the wave function of the
$q\bar{q}$ fluctuation of the photon, which at high energy $\nu$
is formed at a large distance (the coherence length)
\beq
l_{c}={{2\nu}/(Q^{2}+m_{V}^{2})}
\label{eq:2.2.2}
\endeq
in front of the target nucleon. The most important feature of
$\Psi_{\gamma^{*}}(r,z)$ as derived in [15]
is an exponential decrease
at large size
\beq
\Psi_{\gamma^{*}}(r,z) \propto \
\exp(-\varepsilon r)  \, ,
\label{eq:2.2.3}
\endeq
where
\beq
\varepsilon^{2} = m_{q}^{2}+z(1-z)Q^{2}
\label{eq:2.2.4}
\endeq
and  $m_{q}$ is the quark mass. Therefore, the amplitude
(\ref{eq:2.2.1}) recives the dominant contribution from
$r\sim r_{S} \approx 3/\varepsilon$. In the nonrelativistic
quarkonium $z \approx {1\over 2}$, $m_{V}\approx 2m_{q}$, and we
conclude that the vector meson production amplitude probes
the dipole cross section at the scanning radius $r_{S}$
given by [9,10,23]
\beq
r_{S} = { 6 \over \sqrt{m_{V}^{2}+Q^{2}}} \, .
\label{eq:2.2.5}
\endeq
The scanning phenomenon is qualitatively illustrated in Fig.~2.
If the scanning radius $r_{S} \lsim R_{V}$, where $R_{V}$ is the
radius of the vector meson, then for the (T) transverse and (L)
longitudinal mesons
\beq
 M_{T} \propto r_{S}^{2}\sigma(r_{S}) \propto
 {1\over (Q^{2}+m_{V}^{2})^{2} }
\label{eq:2.2.6}
\endeq
\beq
 M_{L} \approx {\sqrt{Q^{2}}\over m_{V}}M_{T}
 \propto
{\sqrt{Q^{2}}\over m_{V}}
 {1\over (Q^{2}+m_{V}^{2})^{2}}
\label{eq:2.2.7}
\endeq
This prediction [10] of the dominance
of the longitudinal cross section
agrees with the E665 [5] and the NMC [25] data.
Compared to the vector dominance model, the predicted amplitudes
contain extra $(Q^{2}+m_{V}^{2})$ in the denominator, the
origin of which is precisely in
CT property $\sigma(r_{S})\propto r_{S}^{2}$.

The scanning radius $r_{S}$ decreases, and the virtual photon shrinks,
with increasing $Q^{2}$. Notice, however, the large numerical factor in
Eq.~(\ref{eq:2.2.5}), for which the onset of small-size dominance
requires very large $Q^{2}$. Remarkably, this large factor derives
precisely from CT property of the dipole cross section. Because of
this large scanning radius, the vector meson production probes
the gluon structure function $G(x,q^{2})$ at [23]
\beq
q^{2} \sim 0.2 (Q^{2}+m_{V}^{2})
\label{eq:2.2.8}
\endeq
The emergence of this very low factorization scale was not noticed
in [26].

Because of a large scanning radius, the simple nonrelativistic
approximation remains viable in quite a broad range of $Q^{2}$,
and the production amplitude can be calculated using the
constituent quark wave functions of vector mesons. For the same
reason of large $r_{S}$, the asymptotic predictions [10]
$\sigma_{T}
\propto 1/Q^{8}$ and $\sigma_{L}  \propto 1/Q^{6}$, can not
readily be tested at $Q^{2}\lsim (10-20)$GeV$^{2}$ studied sofar.

In the conventional quark model, the radii of the $\rho$ and $\pi$
mesons are about identical. Once the radius of the $\rho^{0}$ is
specified, further predictions [23,27] for the $\rho^{0}$ production
cross section
do not contain any adjustable parameters. They are presented
in Fig.~3. We use the low-energy dipole cross section of Ref.~[15].
Shown is the combination of the longitudinal and transverse cross
sections as measured by the NMC collaboration.
The agreement with the recent NMC data [25] is excellent.
This agreement
in a broad range of $Q^{2}$ shows we have a
good understanding of the dipole cross section from the hadronic
scale $r \sim $(1-2)f down to the smallest dipole size $r\sim 0.3$f
achieved in the NMC experiment at $Q^{2}\sim20$GeV$^{2}$. Over this
range of radii $r$, the dipole cross section drops by approximately
one order in magnitude, and this agreement of the total production
rate with experiment is by itself a very important test of CT.
Similar calculations [23] give an excellent
description of the $J/\Psi$ production and of the real
photoproduction of the $\rho^{0}$ at HERA.



\subsection{CT in exclusive production on nuclei}

Having tested CT property of the dipole cross section in the
production on free nucleons, now we turn our attention to the
production on nuclei.
The $q\bar{q}$ pair produced on the target nucleon, recombines
into the observed vector meson with the recombination (formation)
length
\beq
l_{f}={\nu\over  m_{V}\Delta m}\,,
\label{eq:2.3.1}
\endeq
where $\Delta m $ is the typical level splitting in the
quarkonium. At high energy $\nu$ both $l_{c}$ and $l_{f}$
exceed the radius $R_{A}$ of the target nucleus, which greatly
simplifies the theoretical analysis. The E665
data [5]  correspond to this situation. In this high-energy
limit,
nuclear transparency for the incoherent (quasielastic) production
on a nucleus $\gamma^{*}\, A \rightarrow V \,A^{*}$
equals [12,22] (we suppress the energy-dpendence of
$\sigma(\nu,r)$)
\arr
T_{A}={\sigma_{A} \over A\sigma_{p}}={1\over A}
\int d^{2}\vec{b} T(b)
{\langle V |\sigma(r)
\exp\left[-{1\over 2} \sigma(r)T(b)\right] |\gamma^{*}
\rangle^{2} \over
\langle V|\sigma(r)|\gamma^{*}\rangle^{2} } \nonumber\\
=
1-\Sigma_{V} {1\over A}\int d^{2}\vec{b}T(b)^{2} +... \,\, ,
\label{eq:2.3.2}
\endarr
where
\beq
T(b)=\int dz n_{A}(b,z)
\label{eq:2.3.3}
\endeq
is the optical thickness
of a nucleus at the impact parameter
$b$ and $n_{A}(b,z)$ is the
nuclear matter density. In the quasielastic
production one sums over all excitations and breakup of the
target nucleus.
The total cross section of the coherent (elastic)
production $\gamma^{*}\, A \rightarrow V\,A$, when the target
nucleus remains in the ground state,
equals [22]
\arr
\sigma_{coh}(VA) =
4\int d^{2}\vec{b}
\left|\langle V |1-
\exp\left[-{1\over 2} \sigma(r)T(b)\right] |\gamma^{*     }
\rangle\right|^{2}\nonumber\\ =
16\pi\left.
{d\sigma(\gamma^{*}N \rightarrow VN)\over dt}\right|_{t=0}
\int d^{2}\vec{b}
T(b)^{2}\left[1-
{1\over 2}\Sigma_{V}T(b) +... \right]\, .
\label{eq:2.3.4}
\endarr
In Eqs.~(\ref{eq:2.3.2},\ref{eq:2.3.4}) we have explicitly shown
the leading terms of final state interaction
(FSI). The strength of FSI is measured by the
observable [12]
\beq
\Sigma_{V}={\langle V|\sigma(r)^{2}|\gamma^{*}\rangle
\over \langle V|\sigma(r)|\gamma^{*}\rangle } \,.
\label{eq:2.3.5}
\end{equation}
Evidently, the matrix element in the numerator of (\ref{eq:2.3.4})
is dominated by
\beq
r\sim r_{FSI}={5\over 3}r_{S}= {10 \sqrt{Q^{2}+m_{V}^{2}}} \, ,
\label{eq:2.3.6}
\endeq
which gives an estimate [10,23]
\beq
\Sigma_{V} \approx \sigma(r_{FSI})
\label{eq:2.3.7}
\endeq
CT and/or weak FSI set in when $r_{FSI} \ll R_{V}$, i.e., when
$\Sigma_{V} \ll
\sigma_{tot}(VN) $.
In this regime of CT, the  $\Sigma_{V}$ is insensitive to
the wave function of the vector meson, so that
predictions of FSI effects are less model independent than
predictions for the total production cross section.
The large value of $r_{FSI}$ Eq.~(\ref{eq:2.3.6}) implies that
FSI only slowly vanishes with the increase of $Q^{2}$, and this
slow onset of CT is driven by the very mechanism of CT.



\subsection{The E665 experiment [5]: the decisive proof of CT}

Our predictions [9,10] for nuclear effects in the coherent and
incoherent exclusive $\rho^{0}$ production are compared with
the E665 data in Figs.~4-7. In Fig.~4 we show
nuclear transparency for the incoherent production.
Nuclear attenuation
is very strong at small $Q^{2}$ and gradually decreases with
$Q^{2}$. The effect
is particularly dramatic for the heavy
nuclei ($Ca,\, Pb$), and leaves no doubts the E665 observed the
onset of CT. The predicted $Q^{2}$ dependence
of nuclear transparency for
the forward coherent production on nuclei
$
T_{A}^{(coh)} = ({d\sigma_{A}^{(coh)}/A^{2} d\sigma_{N}}
)|_{t=0}$ is shown in Fig.~5.
We predict a rise of $T_{A}^{(coh)}$ with $Q^{2}$ towards
$T_{A}^{(coh)}=1$ for vanishing FSI.
The predicted $Q^{2}$ dependence
of the coherent production cross section relative
to the cross section for the carbon nucleus is presented in Fig.~6.
In the regime of vanishing FSI,
\beq
R_{coh}^{(CT)}(A/C)= {12 \sigma_{A} \over A\sigma_{C}}
\approx {AR_{ch}(C)^{2} \over 12R_{ch}(A)^{2}}\, ,
\label{eq:2.4.1}
\endeq
which gives $R_{coh}^{(CT)}(Ca/C)= 1.56$ and
$R_{coh}^{(CT)}(Pb/C) = 3.25$.
Here $R_{ch}(A)$ is
the charge radius of a nucleus.
The observed growth of the
$Pb/C$ ratio with increasing $Q^{2}$ gives
a solid evidence for the onset of CT.

The (approximate) $A^{\alpha}$ parametrization is a convenient
short-hand representation of the $A$-dependence of nuclear
cross sections, although the so-defined exponent $\alpha$ slightly
depends on
the range of the mass number $A$ used in the fit.
Then, Eq.~(\ref{eq:2.3.2}) predicts that $\alpha_{inc}(Q^{2})$ tends
to 1 from below, as $Q^{2}$ increases. In the limit of vanishing
FSI Eqs.~(\ref{eq:2.3.4},\ref{eq:2.4.1}) predict
$\sigma_{coh} \sim    A^{4/3}$, so
that $\alpha_{coh}(Q^{2})$ tends to $\approx {4\over 3}$
from below as $Q^{2}$ increases
(more accurate analysis shows that the no-FSI cross
section in the $C-Pb$ range of nuclei has the exponent
$\alpha_{coh}
\approx 1.39$).
The agreement between the theory and the E665 fits is good (Fig.~7).
Both the $\alpha_{coh}(Q^{2})$ and
$\alpha_{inc}(Q^{2})$ rise with $Q^{2}$, which is still another
way of stating that the E665 data confirm the onset of CT.
Even at the highest $Q^{2} \sim 10$ GeV$^{2}$ of the E665 experiment,
the residual nuclear attenuation is still strong, confirming
our prediction of large value of $r_{FSI}$ Eq.~(\ref{eq:2.3.6}).


\subsection{Scaling properties of nuclear attenuation:
electroproduction of the $\rho^{0}$ {\it vs.} the real
photoproduction of the $J/\Psi$}

Eq.~(\ref{eq:2.3.6}) in conjunction with Eq.~(\ref{eq:2.2.5})
suggests that nuclear attenuation approximately scales with
$(Q^{2}+m_{V}^{2})$. Specifically, we predict identical nuclear
transparency in the real photoproduction ($Q^{2}=0$) of the
$J/\Psi$ and electroproduction of the $\rho^{0}$ at
$Q^{2}\approx m_{J/\Psi}^{2}-m_{\rho}^{2}$, which nicely agrees
with the experiment:

For the $\rho^{0}$ production
at $\langle Q^{2} \rangle =7$ GeV$^{2}$ the
E665 experiment gives $T_{Pb}/T_{C} = 0.6 \pm 0.25$.
This can be
compared with the NMC result $T_{Sn}/T_{C}= 0.7 \pm 0.1$ for
the real photoproduction of $J/\Psi$ in the similar energy
range [28].
Similar scaling law holds for the coherent production of
the $\rho^{0}$ and the $J/\Psi$.
In the regime of vanishing FSI
Eq.~(\ref{eq:2.3.2}) gives [10,22]
$R_{coh}(Sn/C)=2.76$, $R_{coh}(Fe/Be)=2.82$, $R_{coh}(Pb/Be)=
4.79$. The experimental data on the real photoproduction
of $J/\Psi$ give
$R_{coh}(Sn/C)=2.15\pm 0.10$ in the NMC experiment [29] and
$R_{coh}(Fe/Be)=2.28\pm 0.32$, $R_{coh}(Pb/Be)=3.47 \pm 0.50$
in the Fermilab E691 experiment [30]. These data have
successfully been described [22] in the discussed framework
using the dipole cross section [15].
In all cases the observed
$\sim 30\%$ departure
of the observed ratios for the $J/\Psi$ from predictions for
vanishing FSI
is of the same magnitude as in the highest $Q^{2}$
bin of the E665 data on the $\rho^{0}$ production (Fig.~6).
This scaling relationship between production of different
vector mesons [10] provides a very important cross check of the
mechanism of CT.

Regarding the statistical accuracy of the data, the potential
of the FNAL and CERN muon scattering experiments is nearly
exhausted. Now we shall discuss what new can be done at CEBAF
assuming the energy range $\nu \lsim (8-10)$GeV and
$Q^{2} \lsim (4-6)$GeV$^{2}$.



\section{What next: Vector mesons at CEBAF}


\subsection{Two scales: the coherence length and the formation
length}

The coherence length $l_{c}$ and the formation length $l_{f}$
control the two different aspects of nuclear attenuation.
The formation length tells how rapidly the electropoduced
$q\bar{q}$ state evolves into the observed hadron. For the
$\rho^{0}$ production,
\beq
l_{f} \sim 0.4{\rm f}\cdot\left({\nu \over 1 {\rm GeV}}\right)
  \, .
\label{eq:3.1.1}
\endeq
In the low energy limit of $l_{f}\ll R_{A}$  we have an
instantaneous formation of the final-state hadron, which
then attenuates with the free-nucleon cross section and CT
effects are absent. At high energies $l_{f} > R_{A}$,
the formation of the observed hadron takes place behind the
nucleus, and CT becomes possible. For the observation of the
fully developed CT one needs
\beq
\nu \gsim (3-4)\cdot A^{1/3}\,{\rm GeV} \, .
\label{eq:3.1.2}
\endeq
CT effects already
start showing up, though, if $l_{f} \gsim l_{int}$,
where the interaction length, or the mean free path, equals
\beq
l_{int} = {1 \over n_{A}\sigma_{tot}(VN) } \approx
2{\rm f}\cdot\left( {30 mb \over \sigma_{tot}(VN) } \right)\, .
\label{eq:3.1.3}
\endeq
Therefore, purely kinematically, CT effects are within the reach
of CEBAF experiments after the $8-12$GeV energy upgrade. Notice,
that the formation length $l_{f}$ does not depend on the photon's
virtuality $Q^{2}$.

On the other hand, the coherence length $l_{c}$ tells at which
distance from the absorption point the pointlike photon becomes
the hadronlike $q\bar{q}$ pair. If $l_{c} \gsim R_{A}$, then
the whole thickness of the nucleus contributes to attenuation
of the $q\bar{q}$ pair. Changing the virtuality of the photon
$Q^{2}$ and/or reducing the photon's energy $\nu$, one can make
$l_{c} \ll R_{A}$. In this case, the incoming photon is absorbed
approximately uniformly over the volume of the target nucleus,
and attenuation of the produced $q\bar{q}$ pair will take place
over half of the total thickness of the nucleus. In the
practically interesting  cases of $Q^{2}\gg m_{V}^{2}$ for the
light mesons, or in the real and virtual photoproduction of
heavy quarkonia $J/\Psi,\,\Upsilon$ we have $l_{c} \ll l_{f}$.
If $l_{f} \gsim R_{A}$, but $l_{c}\ll R_{A}$, then nuclear
transparency equals
\beq
T_{A}={1 \over A}\int d^{2}\vec{b}dz n_{A}(b,z)
{ \langle V|\sigma(\rho)\exp[-{1 \over 2}\sigma(\rho)t(b,z)]
|\gamma^{*}\rangle ^{2}
\over
\langle V|\sigma(\rho)|\gamma^{*}\rangle^{2} }
\label{eq:3.1.4}
\endeq
where $t(b,z)$ is the partial optical thickness of the nucleus
\beq
t(b,z)=\int_{z}^{\infty}dz'n_{A}(b,z')\, .
\label{eq:3.1.5}
\endeq
Notice, that $t(b,z)  < T(b)$, and compared to the high-energy
limit (\ref{eq:2.3.2}), here the attenuation effect in the nuclear
matrix element is weaker. Consequently, a
correct treatment of the coherency
effects leads to a prediction [9,12,21,22] of the increase of nuclear
attenuation with increasing energy $\nu$:
\beq
T_{A}(l_{c} \gsim R_{A}) < T_{A}(l_{c} \ll R_{A})
\label{eq:3.1.6}
\endeq
The energy dependence of the nuclear transparency is given by an
approximate interpolation formula
\beq
T_{A}(\nu)\approx T_{A}(l{c} << R_{A})+G_{A}(\kappa)^{2}
[Tr_{A}(l_{c} >R_{A})-Tr_{A}(l_{c} <<R_{A})]
\label{eq:3.1.7}
\endeq
where $G_{A}(\kappa)$ is the charge form factor of the target
nucleus and $\kappa$ is the longitudinal momentum transfer in
the transition $\gamma^{*} N\rightarrow VN$:
\beq
\kappa ={1\over l_{c}}={Q^{2}+m_{V}^{2}\over 2\nu}\,.
\label{eq:3.1.8}
\endeq
The predicted energy dependence of nuclear transparency
 [22] is in excellent agreement with the NMC
data on the
real photoproduction of the $J/\Psi$ [28] shown in Fig.~8.
The NMC data rule out
the widely discussed models of "quantum" and "classical" diffusion
by Farrar, Frankfurt and Strikman [31], which has also predicted
the precocious CT in the $A(e,e'p)$ scattering (for the review
see [6]).



\subsection{ Quantum evolution and energy dependence of FSI.}

If $l_{f} < R_{A}$, the frozen-size approximation is no longer
applicable, and spatial expansion of the $q\bar{q}$ pair
during its propagation inside a nucleus
becomes important. The consistent path-intergal description of
the spatial expansion effects was developed in [21]. Interaction
of the $q\bar{q}$ pair with the nuclear matter can be described
by the optical potential $V_{opt}(r) \propto \sigma(r)n_{A}(b,z)$,
which adds to the confining $q\bar{q}$ potential $U(r)$. The
longitudinal distance $z$ plays the role of time. The path
integral technique provides a systematic procedure for calculating
the evolution kernel ${\cal K}(r',r,t)$ in the combined potential
$U(r)+V_{opt}(r)$. The particularly elegant exact solution for the
evolution kernel is found  for the harmonic oscillator
$U(r)\propto r^{2}$ and $\sigma(r) \propto r^{2}$ [21].

The resulting predictions for the energy dependence of nuclear
transparency $T_{A}$ for the production of different vector
mesons are shown in Figs.~9,10.  The salient features of
$T_{A}$ are [9,21,22]:
\begin{itemize}

\item
At small $Q^{2}$, nuclear transparency for the $\rho^{0}$ and
the $J/\Psi$ decreases with energy $\nu$, starting at the value
given by Eq.~(\ref{eq:3.1.4}) and levelling off at the
value given by Eq.~(\ref{eq:2.3.2}). This decrease is
due to the increase of the coherence length $l_{c}$, discussed
in section 3.1.

\item
At larger $Q^{2}$, the trend changes: $T_{A}$ first increases
with $Q^{2}$, and then decreases for the same reason of the
rise of
the coherence length $l_{c}$. In agreement with
Eqs.~(\ref{eq:3.1.7},\ref{eq:3.1.8}),
the larger is $Q^{2}$, the higher is the energy $\nu$ at which
the levelling off of $T_{A}$ takes place.

\item
For the $\Upsilon$ production, nuclear transparency $T_{A}$
starts increasing with energy at all values of $Q^{2}$, in
close similarity to the $J/\Psi$ production at large $Q^{2}$,
and in agreement with our conclusion that nuclear attenuation
scales with $(Q^{2}+m_{V}^{2})$, see section 2.5. For instance,
we predict $T_{A}(J/\Psi, Q^{2}=100{\rm GeV}^{2}) \approx
T_{A}(\Upsilon, Q^{2}=0)$.

\item
The radial excitations $\Psi',\,\Upsilon'$ have larger size,
and larger free-nucleon cross section thereof, than the ground
states $J/\Psi$ and $\Upsilon$, respectively. Nonetheless,
the radial excitations are predicted to have weaker nuclear
attenuation, which is a completely counterintuitive result.

\item
At last but not the least, for the $\rho^{0}$ production we
predict very rich pattern of the $\nu$ and $Q^{2}$ dependence
precisely in the kinematical range of CEBAF. CT effects are
large and can readily be observed at CEBAF.
The $\rho'$ production
will be treated in section 3.7.

\end{itemize}

These properties of nuclear transparency can best be understood
in terms of the interplay of CT with the node effect.



\subsection{CT and the node effect: antishadowing phenomenon.}

The wave function of the radial excitation $V'(2S)$ has a node.
For this reason, in the $V'$ production amplituide there is the
node efffect -  cancellations between the contributions from $r$
below, and above, the node. The product
$\sigma(r)\Psi_{\gamma^{*}}(z,r)$ acts as a distribution,
which probes the wave function of the $V(1S)$ and $V(2S)$
states at the scanning radius $\sim r_{S}$ [9,10,12]
 and the node
effect evidently depends on the scanning radius $r_{S}$,
see Fig.~2.

If the node effect is strong, even the slight variations of
$r_{S}$ lead to an anomalously rapid variation of the $V'(2S)$
production amplitude, which must be contrasted to the smooth
$Q^{2}$ and $r_{S}$ dependence of the $V(1S)$ production amplitude.
Evidently, the stronger is the node effect and the smaller is the
$V'(2S)$ production amplitude, the higher is the sensitivity to
the model for the $V'(2S)$ wave function, and in some cases only
firm statement will be the fact of the strong suppression of the
$V'(2S)$ production.

For the real photoproduction of the $\Psi'$, the calculations in
[21] gave $\sigma(\gamma N\rightarrow \Psi' N)/
\sigma(\gamma N\rightarrow J/\Psi N)=0.17 $, which is in excellent
agreement with the NMC result $0.20 \pm 0.05(stat) \pm 0.07(syst)$
for this ratio [29]. In this case the node effect is already rather
strong for the fact that the scanning radius $r_{S}(Q^{2}=0)$
is rather close to the $J/\Psi$ radius $R_{J/\Psi}$.
For the $\Upsilon'$, the scanning radius is substantially smaller
than $R_{\Upsilon}$, which is realtively large for the small strong
coupling $\alpha_{S}(R_{\Upsilon})$. For the light mesons, the
scanning radius $r_{S}$ is larger and, at small $Q^{2}$, the node
effect is much stronger, see section 3.7.

Because $r_{FSI}$ is larger than $r_{S}$, the node effect
in the strength of FSI
given by Eq.~(\ref{eq:2.3.5}) becomes stronger.
For the $J/\Psi$, one finds the {\it \bf overcompensation}:
$\langle V|\sigma(r)^{2}|\gamma\rangle < 0$, which leads to
$\Sigma_{V} < 0$ and to the antishadowing phenomenon $T_{A} > 1$
shown in Fig.~10 .
For the $\Upsilon'$, we find the {\it \bf undercompensation}:
$\langle V|\sigma(r)^{2}|\gamma\rangle > 0$, which leads to
$\Sigma_{V} > 0$ and to the shadowing $T_{A} < 1$. None the less,
the node effect shows up: for the $\Upsilon'$ with its large
radius, nuclear attenuation is weaker than for the $\Upsilon$ !
With increasing $Q^{2}$, when $r_{FSI} \ll R_{V}$, the node effect
becomes negligible, and the $V(1S)$ and $V'(2S)$ states will
have identical nuclear attenuation, see Fig.~10.



\subsection{The interplay of CT, of the node effect and of quantum
evolution}

Of course, the $\Psi'$ has a larger radius and larger free-nucleon
cross section $\sigma_{tot}(\Psi' N) \sim$(2.5-3)$
\sigma_{tot}(J/\Psi\, N)$. How come, then, that
in the real photoproduction $\Sigma_{\Psi'}
< 0$ and we find the antishadowing of the strongly interacting
$\Psi'$ alongside with shadowing for the $J/\Psi$?

Although the above derivation of antishadowing was (deceptively)
simple,
still another look at antishadowing and
the variation of energy dependence of nuclear
transparency with $Q^{2}$ is in order [9]. Let us consider for
simplicity the $J/\Psi,\,\Psi'$ system. The numerator of the
strength of FSI~ $\Sigma_{V}$ can be expanded in terms of
the complete set of intermediate states $|i\rangle>$ of charmonium
\beq
\langle V|\sigma(\rho)^{2}|\gamma^{*}\rangle =
\sum _{i} \langle V|\sigma(\rho)|V_{i}\rangle
\langle V_{i} |\sigma(\rho)|\gamma^{*}\rangle \, .
\label{eq.3.4.1}
\endeq
In terms of this expansion, antishadowing of the photoproduction
of the $\Psi'$ comes from the destructive interference of the two
dominant intermediate states: the direct, VDM-like rescattering
\beq
\gamma^{*} \rightarrow \Psi' \rightarrow \Psi'
\label{eq.3.4.2}
\endeq
and the off-diagonal rescattering
\beq
\gamma^{*} \rightarrow J/\Psi \rightarrow \Psi'
\label{eq:3.4.3}
\endeq
(there is a small contribution from other intermediate states too).

Then,
for the $\Psi'$ production, the strength of FSI is given by
\beq
\Sigma_{\Psi'}=\sigma_{tot}(\Psi' N)+
M(J/\Psi\, N\rightarrow \Psi' N)\cdot{
M(\gamma^{*} N \rightarrow J/\Psi\,N) \over
M(\gamma^{*} N\rightarrow \Psi'N)} \, .
\label{eq:3.4.4}
\endeq

Because of the interplay of CT and the node effect, we have
$M(\gamma^{*} N \rightarrow J/\Psi\,N)/
M(\gamma^{*} N\rightarrow \Psi'N) \gg 1$. For the same interplay
of CT and the node effect, there is an {\sl overcompensation} in
the amplitude of the off-diagonal transition
\beq
M(J/\Psi\, N\rightarrow \Psi' N) < 0\, ,
\label{eq:3.4.5}
\endeq
and numerically this amplitude is not very small,
$M(J/\Psi\, N\rightarrow \Psi' N) \sim -\sigma_{tot}(J/\Psi\,N)$.
Consequently,
the second, negative valued, off-diagonal term in $\Sigma_{\Psi'}$
takes over the $\sigma_{tot}(\Psi' N)$ (the higher intermediate
states also slightly contribute to the antishadowing effect).

What happens with increasing $Q^{2}$ is very simple:
The scanning radius $r_{S}$ decreases with $Q^{2}$ and the
node effect become weaker, the ratio of amplitudes
$M(\gamma^{*} N \rightarrow J/\Psi\,N)/
M(\gamma^{*} N\rightarrow \Psi'N)$ decreases with $Q^{2}$ and
tends to approximately unity, whereas $\sigma_{tot}(\Psi' N)$
and $M(J/\Psi\, N\rightarrow \Psi' N)$ do not depend on $Q^{2}$.
As a result, the off-diagonal contribution in Eq.~(\ref{eq:3.4.4})
becomes small, and at large $Q^{2}$ the antishadowing of $\Psi'$
changes to the shadowing.

Similarly, for the $J/\Psi$ production
\beq
\Sigma_{J/\Psi}=\sigma_{tot}(J/\Psi\, N)+
M(J/\Psi\, N\rightarrow \Psi' N)\cdot{
M(\gamma^{*} N \rightarrow \Psi'N) \over
M(\gamma^{*} N\rightarrow J/\Psi\,N)} \, .
\label{eq:3.4.6}
\endeq
In this case, for the real photoproduction the off-diagonal
contribution is small because of
$M(\gamma^{*} N \rightarrow J/\Psi\,N)/
M(\gamma^{*} N\rightarrow \Psi'N) \gg 1$, which is a consequence
of CT. Therefore, for this numerical reason, for the real
photoproduction we find
\beq
\Sigma_{V}\approx \sigma_{tot}(J/\Psi\, N)
\label{eq:3.4.7}
\endeq
which explains why the predicted nuclear attenuation of the
$J/\Psi$ is
marginaly similar [21,22] to the vector dominance model estimates.
With increasing $Q^{2}$, the off-diagonal term in (\ref{eq:3.4.6})
increases and starts camcelling the term $\sigma_{tot}(J/\Psi\,N)$,
thus depleting $\Sigma_{J/\Psi}$  and leading to CT effect.

Consider now the $J/\Psi$ and $\Psi'$ photoproduction at a finite
energy. The intermediate $J/\Psi$ and $\Psi'$ have different masses
$m_{1}$ and $m_{2}$ and propagate with momenta which differ by
\beq
\kappa_{21} = {m_{2}^{2}-m_{1}^{2}\over 2\nu} \approx {1\over l_{f}}
\label{eq:3.4.8}
\endeq
If the production and rescattering points are a distance
$\Delta z = z_{2}-z_{1}$ apart, then the off-diagonal
contribution  to the rescattering amplitude acquires the
phase factor $\exp[i\kappa_{21}(z_{2}-z_{1})]$ with respect
to the elastic rescattering contribution [32,33].
Upon the integration over $z_{1,2}$, the effect of this
phase factor is
\beq
 \left< \exp[i\kappa_{21}(z_{2}-z_{1})] \right> =
G_{A}^{2}(\kappa_{21})  \, ,
\label{eq:3.4.9}
\endeq
and the off-diagonal
contributions to $\Sigma_{V}$ will enter with the suppression
factor $G_{A}(\kappa_{i1})^{2}$ (for the detailed analysis see
[7,8,9,13])
\beq
\Sigma_{\Psi'}=\sigma_{tot}(\Psi' N)+
M(J/\Psi\, N\rightarrow \Psi' N)\cdot{
M(\gamma^{*} N \rightarrow J/\Psi\,N) \over
M(\gamma^{*} N\rightarrow \Psi'N)}
G_{A}^{2}(\kappa_{12})
\, .
\label{eq:3.4.10}
\endeq
\beq
\Sigma_{J/\Psi}=\sigma_{tot}(J/\Psi\, N)+
M(J/\Psi\, N\rightarrow \Psi' N)\cdot{
M(\gamma^{*} N \rightarrow \Psi'N) \over
M(\gamma^{*} N\rightarrow J/\Psi\,N)}
G_{A}^{2}(\kappa_{12})
\, .
\label{eq:3.4.11}
\endeq
Evidently, at low energy such that $\kappa_{12}R_{A} \gsim 1$,
i.e., at $l_{f} \ll R_{A}$, the nuclear form factor vanishes
$G_{A}(\kappa_{12})^{2} \ll 1$. Only the diagonal contributions
to $\Sigma_{V}$ survive, and CT effects which come from the
off-diagonal terms in $\Sigma_{V}$, vanish at low energy.
For instance, for the $\Psi'$  one starts with the shadowing
$T_{A} < 1$, which with increasing energy and the opening of the
off-diagonal channels, changes to the antishadowing.
For the $J/\Psi$ and $\rho^{0}$, at small energy there is a
competition of CT effect which rises with energy, and of the effect
of growth of the coherence length. The latter takes over at small
$Q^{2}$, whereas at larger $Q^{2}$ the opening of the off-diagonal
transitions leads to a rapid near-threshold rise of nuclear
transparency. For the $\rho^{0}$ production, nuclear transparency
$T_{A}$ is a lively function of energy $\nu$ and $Q^{2}$ in
precisely the kinematical region accessible at CEBAF.



\subsection{Measuring the $J/\Psi$-nucleon cross section at
CEBAF}

The smallness of the off-diagonal rescattering in the real
photoproduction of the $J/\Psi$, see Eqs.~(\ref{eq:3.4.7}),
(\ref{eq:3.4.11}),
leads to an important prediction [9,22] that nuclear attenuation
allows to evaluate  $\sigma_{tot}(J/\Psi \,N)$ using the
conventional VDM formulas. This suggestion was carried
over in an analysis [34] of the data [29,30] on the coherent
photoproduction on nuclei with the result
$\sigma_{tot}(J/\Psi\, N)
\sim $(5-7)mb. The CEBAF experiments will allow measurement of
this cross section at low energy, although being very close to
the threshold requires a good understanding of the Fermi-smearing
effects.

For the $\Psi'$ photoproduction at low energies, the off-diagonal
transitions (\ref{eq:3.4.3})
are non-negligible even close to the threshold,
and for the $\Psi'$ the VDM prediction for nuclear shadowing
breaks down completely: $T_{A}$ is always larger than the VDM
prediction.


\subsection{The coherence and formation lengths revisited:
Vanishing nuclear shadowing in inclusive DIS at CEBAF coexists
with lots of CT in exlusive production at CEBAF}

Above we have repeatedly emphasized that the onset of CT is
entirely controlled by the formation length $l_{f}$, which
does not depend on $Q^{2}$. The onset of CT is quantified
by Eqs.~(\ref{eq:3.4.10},\ref{eq:3.4.11}), in which the formation
length enters via the nuclear form factor $G_{A}(\kappa_{12})$,
where $\kappa_{12}$ is the longitudinal momentum transfer in the
off-diagonal transition $V_{1}N\rightarrow V_{2}N$.

At large $Q^{2}$ we have $l_{c}\ll l_{f}$ and much larger
longitudinal momentum transfer $\kappa$ Eq.~(\ref{eq:3.1.8})
in the transition $\gamma^{*}N \rightarrow VN$. However,
because $\kappa \gg \kappa_{12}$, this momentum transfer is
approximately the same for all intermediate states. At
$l_{c} \ll R_{A}$, the corresponding overall phase factor
simply drops out from the incoherent production cross section.
Neither does this phase factor affect $T_{A}$ dramatically in
the transient regime of $l_{c}\sim R_{A}$, see Eq.~(\ref{eq:3.1.7}).
In the coherent production on nuclei, the major effect of the
momentum tranfer $\kappa$ is that the nuclear production
amplitude acquires the overall factor $G_{A}(\kappa)$, which
significantly suppresses the coherent production amplitude
but does not affect the nuclear attenuation properties.

As we discussed in section 2.2, the vector meson production
on the free nucleon probes the gluon distribution in the
target proton. Stretching the so-called
factorization theorems, one can be tempted to conclude that
nuclear attenuation in the production of vector mesons is
given by the nuclear shadowing of gluon structure function [26].
Indeed, exclusive electroproduction of vector mesons is the
typical diffraction dissociation of the photon
$\gamma^{*}N\rightarrow XN$, and virtual diffractive transitions
$\gamma^{*} \rightarrow X \rightarrow $ in the nuclear forward
Compton scattering amplitude are precisley the source of nuclear
shadowing [15]. But, subtle is the nuclear shadowing!

The mere definition of the shadowed nuclear parton distributions
is only useful provided that the shadowing term by itself
satisfies the conventional evolution equations. To a certain
approximation, this is the case [17]. However, there are no
theorems on the universality of these shadowing corrections
in all hard scattering processes and shadowing corrections
may defy the factorization theorems [16]. Here we present
simple arguments, essentially of kinematical origin, why
the factorzation theorems must be taken with the grain of salt.

The contribution of the intermediate state $X$ to the nuclear
Compton scattering amplitude contains the excitation
$\gamma^{*}N\rightarrow XN$ on one nucleon and the de-excitation
$XN\rightarrow \gamma^{*}N$ on another nucleon. In both
transitions there is a longitidunal momentum transfer $\kappa$
Eq.~(\ref{eq:3.1.8}). Consequently, the corresponding
contribution to nuclear shadowing in the structure function
will enter with the suppression factor $G_{A}(\kappa)^{2}$. The
onset of nuclear shadowing requires $l_{c} \gsim R_{A}$, so that
the larger is $Q^{2}$, the higher energy is required for the
onset of nucler shadowing. In the opposite to that, nuclear
attenuation and CT effects in the nuclear electroproduction of
vector mesons only requires $l_{f} \gsim l_{int},R_{A}$, and this
condition does not depend on $Q^{2}$ and does not require $l_{c}
\gsim R_{A}$. For instance, there will be {\sl \bf no nuclear shadowing
in the inclusive electroproduction on nuclei} in the kinematical
range of the CEBAF experiments, but {\sl \bf lots of CT effects in the
exclusive production of vector mesons on nuclei} at CEBAF. We
wish to emphasize this simple, but important, point in view of
the opposite claims made to this effect by Frankfurt and Strikman at
this Workshop.


\subsection{Anomalous electroproduction of radial excitations
$\rho',\,\phi'$ : CEBAF's new window at CT }

For the light vector mesons, at small $Q^{2}$ the scanning radius
$r_{S} \sim R_{V}$, and there is an exciting, and most likely,
possibility of the {\it overcompensation} already in the free-nucleon
production amplitude:$M=\langle V' |\sigma(r)|\gamma^{*}\rangle
< 0\, .$ The $\rho'$ production on nuclei is
indispensable for testing the node effect and $Q^{2}$
dependence of the scanning radius $r_{S}$, because nuclear
attenuation gives still another handle on the scanning radius
[9,13,27].
For the sake of simplicity, we discuss the quasielastic
(incoherent) $\rho'$ production on nuclei assuming that
$l_{f} \gsim R_{A}$. Extension to lower energies and to the
coherent production is straightforward and the prediction
[13,27] of
the anomalous $Q^{2}$ and $A$ dependence persists in these
cases too.

The $A$-dependence of the node effect comes from the
nuclear attenuation
$\exp[-{1\over 2} \sigma(r)T(b)]$ in the nuclear matrix
element $M_{A}(T)=
 \langle V |\sigma(r)
\exp\left[-{1\over 2} \sigma(r)T(b)\right] |\gamma^{*}
\rangle$. The possibility of the $A$-dependent node effect
effect in hadronic diffraction
production on nuclei $hA\rightarrow h^{*}A$
was pointed out in [35].

Firstly, consider the $Q^{2}$ dependence of the $\rho'/\rho^{0}$
ratio on the free nucleon. Increasing $Q^{2}$ and decreasing the
scanning radius $r_{S}$, one will bring the $\rho'$ production on
the free nucleon to the exact node effect, and the $\rho'/\rho^{0}$
ratio takes the minimum value at a certain finite $Q^{2}$, see
Fig.~11. Because of the $r$-dependence of the attenuation
factor, in the nuclear amplitude the node effect will
be incomplete. Consequently, as a function of $Q^{2}$, nuclear
transparency $T_{A}$ will have a spike $T_{A} \gg 1$ at a
finite value of $Q^{2}$ [9].

Secondly, consider the $\rho '$ production on nuclei at a
fixed value of $Q^{2}$ such that the
free nucleon amplitude is still in the overcompensation regime.
Increasing $A$ and enhancing the importance of the attenuation
factor $\exp[-{1\over 2} \sigma(r)T(b)]$,
we shall bring the nuclear amplitude
to the nearly exact compensation regime. Therefore, the
$\rho'/\rho^{0}$ production ratio, as well as nuclear transparency
for the $\rho'$ production,  will decrease with $A$ and
take a minimum value at a certain finite
$A$. With the further increase of $A$, the undercompensation regime
takes over, and we encounter very counterintuitive situation:
nuclear transparency for the $\rho'$ is larger for heavier, more
strongly
absorbing nuclei!  This situation is illustrated in Fig.~12a and
must be contrasted with a
smooth and uneventful decrease of transparency
for the $\rho^{0}$ production on heavy nuclei.

With the further increase of $Q^{2}$ one enters the pure
undercompensation regime for all the targets.
Nuclear undoing of the node effect enhances $M_{A}$
and nuclear transparency $T_{A}$, whereas the overall
attenuation factor
$\exp[-{1\over 2}\sigma(r)T(b)]$ decreases $M_{A}$ and
$T_{A}$. Of these
two competing effects, the former remains stronger and we
find antishadowing of the $\rho'$ production in a broad
range of $A$ and $Q^{2}$, see Figs.~12c-12e.
Typically, we find a nuclear enhancement
of the $\rho'/\rho^{0}$ production ratio on heavy targets
by one order in the
magnitude with respect to the free nucleon target. This makes
leptoproduction on nuclei the $\rho'$ factory, and the $\rho'$
production experiments at CEBAF can contribute much to the
poorly understood spectroscopy of the radially excited vector
mesons.
Only at a relatively large
$Q^{2}\gsim 2$GeV$^{2}$, the attenuation effect takes over,
and
nuclear transparency for the $\rho '$ production will
start decreasing monotonically with $A$ (Fig.~12f).
Still, this decrease
is much weaker
than for the $\rho^{0}$ meson.
At very large $Q^{2}$, when the node effect disappears
because of the small scanning radius $r_{S}$, nuclear transparency
for the $\rho^{0}$ and the $\rho '$ production will become
identical. This pattern repeats qualitatively the one studied
for the $J/\Psi$ and $\Psi'$ mesons in sections 3.3 and 3.4 \, .

The above presented results refer to the production of the
transversely polarized $\rho^{0}$ and $\rho'$ mesons. Accurate
separation of the transverse and longitudinal
cross section can easily be done in the high statistics CEBAF
experiments. Here we only wish to mention the interesting
possibility that for the longitudinally polarized $\rho'$ mesons,
the exact node effect is likely to take place at a value of $Q^{2}$
larger than for the transverse $\rho'$, so that polarization of
the produced $\rho'$ can exhibit very rapid change with $Q^{2}$.

The numerical predictions are very sensitive to the position
of the node in the wave function of the $\rho'(2S)$.
It is quite possible that the dip of nuclear transparency $T_{A}$
will take place for targets much heavier than in Figs.~12a,12b,
and
disappearance of the node effect and the onset of the more
conventional nuclear shadowing $T_{A} <1$ for the $\rho '$
production like in Fig.~12f only will take place at much larger
$Q^{2}$. Also, the possibility of the undercompensation at
$Q^{2}=0$ can not be excluded. However, the strikingly different
$A$-dependence of the incoherent $\rho^{0}$ and $\rho'$ production
on nuclei persists in such a broad range of $Q^{2}$ and of the
scanning radius $r_{S}$, that the existence of the phenomenon
of anomalous $A$ and $Q^{2}$ dependence of the $\rho'$ preduction
is not negotiable. It is a direct manifestation of
the color-transparency driven $Q^{2}$ dependence of the
scanning radius and, as such, it deserves a dedicated experimental
study.

The case of the $\phi^{0}$ and $\phi'(1680)$ production was studied
in [24] using the path integral technique [21] modified to
include corrections for the departure from the $\propto r^{2}$
dependence of the dipole cross section $\sigma(r)$. Corrections
to the $\phi^{0}$ production
are negligible, corrections to the $\phi'$
production do not exceed $10\%$. Because of the smaller radius
of the $\phi^{0}, \,\phi'$, the numerical predictions are more
reliable than for the $\rho'$. On the free nucleon,
we find undercompensation for the transverse $\phi'$ production.
There is much
similarity to the $J/\Psi, \,\Psi'$ system, but the node effect
is stronger than for the charmonium.
In the nuclear production of the $\phi'$, the antishadowing
effect shown in Fig.~13 gradually builds up with increasing energy,
and numerically is much stronger than for the
$\Psi'$.
Nuclear attenuation of the $\phi'(1680)$ is weaker than
for the $\phi^{0}$ starting already at low energy (Fig.~14).
Because of the large mass and small radius
of the $\phi^{0}$,
nuclear transparency for the $\phi^{0}$ production
only weakly depends on $Q^{2}$ at CEBAF, but for the $\phi'(1680)$
production the predicted $Q^{2}$ dependence is quite strong
(Fig.~15) and can easily be measured at CEBAF. Here the larger
$Q^{2}$ and smaller scanning radius $r_{S}$ predict increasing
attenuation of the $\phi'(1680)$. However,
for the lead target we expect  the onset
of increasing nuclear transparency at already moderate
$Q^{2} \gsim $1$GeV^{2}$.

Few more comments about the possibilities of CEBAF are worth
while. Because of the strong suppression of the $\rho'/\rho^{0}$
and $\phi'/\phi^{0}$ production ratio by the CT and node effects,
the high luminosity of CEBAF is absolutely crucial for
high-statistics experiments on the $\rho',\,\phi'$ production.
Notice, that the most interesting anomalies in the $A$ and
$Q^{2}$ dependence take place near the minimum of the $\rho'$
production cross section. Furthermore, the observation of the
$\rho'$ production requires detection of its 4-pion decays,
and here one can take advantage of the CLAS multiparticle
spectrometer available at CEBAF.



\section{FSI and nuclear transparency in $A(e,e'p)$ scattering}


\subsection{Multiple-scattering expansion for the nuclear
spectral function}

We are interested in $A(e,e'p)$ scattering at large
$Q^{2} >$(1-2)GeV$^{2}$, when the struck proton has the
kinetic energy $T_{kin} \approx Q^{2}/2m_{p} \gsim $(0.5-1)GeV.
Such a proton has the free nucleon total cross section
$\sigma_{tot}(pN) \approx $40mb. The corresponding mean free
path in the nuclear medium $l_{int} \sim 1.5$f is short and
implies strong FSI and strong nuclear attenuation of struck
protons. At large $Q^{2}$ this FSI is expected to vanish by
virtue of CT.
One needs first a reliable formalism for description of FSI
of the struck proton, and the coupled-channel generalization
[7,8,36,37] of the Glauber's multiple scattering
theory [38] provides the necessary framework.

The quantity measured in the ideal
$A(e,e'p)$ scattering is the spectral
function $S(E_{m},\vec{p}_{m})$ as a function of the missing
energy $E_{m}$ and missing momentum $\vec{p}_{m}=
(p_{m,z},\vec{p}_{\perp})$, for the
kinematics see Fig.~16. The discussion greatly simplifies
if the measured cross section is integrated over the sufficiently
broad range of the missing energy $E_{m}$, when the closure can
be applied [7,8,36]. We assume this is the case.
If the plane-wave impulse approximation (PWIA) were applicable,
then one would have have measured the single-particle
momentum distribution $n_{F}(\vec{p}_{m})$ ([39], for the recent
review see [40]), which is related to the one-body
nuclear density matrix $\rho_{1}(\vec{r},\vec{r}\,')$ as
\arr
d\sigma_{A}
\propto
 n_{F}(\vec{p}_{m})=
{1\over Z}\int dE_{m}S_{PWIA}(E_{m},\vec{p}_{m})=
\int d\vec{r}\,'d\vec{r} \,
\rho_{1}(\vec{r},\vec{r}\,')
\exp[i\vec{p}_{m}(\vec{r}\,'-\vec{r})] \, .
\label{eq:4.1.1}
\endarr
In this case the missing momentum $\vec{p}_{m}$ is precisely the
intranuclear momentum of the struck proton.

PWIA is the theorists dream. The PWIA spectral function is
what theorists do calculate in their models,
 in the real life the outgoing
proton can not be described by the plane wave.
At large $Q^{2} \gsim 1$GeV$^{2}$, FSI of the struck ptoton can
be treated in the Glauber approximation.
The FSI modifies Eq.~(\ref{eq:4.1.1}) [36,37]:
\arr
d\sigma_{A} \propto
{1\over Z}\int dE_{m}S(E_{m},\vec{p}_{m})= ~~~~~~~~~~~~~~~~
\nonumber\\
\int d\vec{r}\,'d\vec{r} \,
\rho_{1}(\vec{r},\vec{r}\,')
\exp[i\vec{p}_{m}(\vec{r}\,'-\vec{r})]
\cdot\exp\left[ t(\vec{b},max(z,z')) \xi(\vec{\Delta}) \right]
\nonumber \\
\cdot \exp\left[
-{1\over 2}(1-i\alpha_{pN})\sigma_{tot}(pN)
t(\vec{b},z)
-{1\over 2}(1+i\alpha_{pN})\sigma_{tot}(pN)
t(\vec{b}',z') \right] \,,
\label{eq:4.1.2}
\endarr
where $\vec{r}=(\vec{b},z),\,\vec{r}\,'=(\vec{b}',z')$,
$\vec{\Delta}=\vec{r}-\vec{r}\,'$, $\alpha_{pN}$
denotes the ratio of the real to imaginary parts of the forward
proton-nucleon scattering amplitude and
\beq
\xi(\vec{\Delta}) =
\int d^2\vec{q} \;\
\frac{d\sigma_{el}(pN)}{d^2\vec{q}} \;\
\exp(i \vec{q}\vec{\Delta}) \, .
\label{eq:4.1.3}
\endeq
The derivation of attenuation factors in (\ref{eq:4.1.2}) uses
the independent particle model. The effect of two-nucleon
correlations in nuclear attenuation was studied in detail in
[41] and found to be negligible.

FSI leads to the four important effects [13,36,37]
\begin{itemize}

\item
Firstly, as a result of the phase factor in the integrand of
(\ref{eq:4.1.2}) which is of the form
$\exp[i{1\over2}\sigma_{tot}(pN)\alpha_{pN}n_{A}(b,z)(z-z')]$
the spectral function  $S(k_{\perp},k_{z})$ is probed
at a shifted value of the longitudinal momentum with
\beq
k_{z}- p_{m,z}=  \Delta p_{m,z}
\sim
{1\over 2}\sigma_{tot}(pN)\alpha_{pN}n_{A} \approx
\alpha_{pN}\cdot 35\, {\rm (MeV/c)} \, .
\label{eq:4.1.4}
\endeq
In the kinematical range of the
NE18 experiment $\alpha_{pN} \sim -0.5$ [42] and $\Delta p_{m,z}
\sim -20$ MeV$/c$ is quite large. This shift makes the measured
$p_{m,z}$ distribution strongly asymmetric about $p_{m,z}=0$.

\item
Secondly, the factor
$\exp\left[ t(\vec{b},z) \xi(\vec{\Delta}) \right]$ in
Eq.~(\ref{eq:4.1.2}) leads to a
broadening of the $p_{\perp}$ distribution. The
multiple elastic-rescattering expansion
for the $(E_{m},p_{m,z})$-integrated spectral
function reads
\beq
f{_A}(\vec{p}_{\perp}) = {1\over Z}\int dE_{m}\,dp_{m,z}\,\,
S(E_{m},p_{m,z},p_{\perp})=
\sum_{\nu = 0}^{\infty}
W^{(\nu)}\,n^{(\nu)}(\vec{p}_{\perp}) \, .
\label{eq:4.1.5}
\endeq
Here
the $p_{\perp}$-distribution in the
$\nu$-fold rescattering $n^{(\nu)}(\vec{p}_{\perp})$ equals
\beq
n^{(\nu)}(\vec{p}_{\perp}) =
\int d^2 \vec{s}  \;\ {B\over \nu\pi}
\exp \left( - {B\over \nu} s^2 \right)
n_{F}(\vec{p}_{\perp} - \vec{s})
\label{eq:4.1.6}
\endeq
where $B$ denotes the diffraction slope for elastic $pN$
scattering, $d\sigma_{el}(pN)/dt \propto \exp(-B|t|)$,
 and
\beq
W^{(\nu)} = \frac{1}{A} \int dz d^2\vec{b} \;\ n_A(\vec{b},z)\,
\exp \left[-\sigma_{tot}(pN) t(\vec{b},z) \right]
{[t(\vec{b},z)
\sigma_{el}(pN)]^{\nu}\over \nu! }
\label{eq:4.1.7}
\endeq
gives the probability of having $\nu$ elastic rescatterings.

\item
Thirdly, FSI introduces the attenuation effect. Namely, whereas
in the PWIA one has $\int d^{3}\vec{p}_{m}\,n_{F}(p_{m})=1$,
with allowance for FSI
\arr
T_{A}= {1\over Z}\int dE_{m}\,d^{3}\vec{p}_{m}\,
S(E_{m},\vec{p}_{m}) = \int d^{2}\vec{p}_{\perp}
\,f_{A}(p_{\perp}) ~~~~~\nonumber\\
=\sum_{\nu=0} W^{(\nu)}
 = \frac{1}{A} \int dz d^2\vec{b} \;\ n_A(\vec{b},z)
\exp \left[-\sigma_{in}(pN) t(\vec{b},z) \right] < 1\, .
\label{eq:4.1.8}
\endarr
In the completely integrated nuclear spectral function
(\ref{eq:4.1.8}), nuclear attenuation is given by $\sigma_{in}(pN)$
[7]. This result is self-obvious: elastic
rescatterings only deflect,
but do not absorb, the struck proton, and the effect of deflection
is not relevant for the full $4\pi$ acceptance.

\item
On the other hand, the forward peak of $f_{A}(p_{\perp})$ at
$\vec{p}_{\perp}=0$  is dominated by
$W^{(0)}$, which is also given by
Eq.~(\ref{eq:4.1.8}) but with $\sigma_{in}(pN)$ substituted by
$\sigma_{tot}(pN)$. In this case elastic rescatterings also
contribute to the observed attenuation.

\end{itemize}

The further discussion will be centered on: (1) how CT affects the
integrated $T_{A}$ and forward (the in-parallel kinematics)
$W^{(0)}$ nuclear transparency; (2) why the onset of CT in
$A(e,e'p)$ scattering is so slow; (3) the theoretical
interpretation of the nonobservation of CT in the NE18
experiment; (4) the discussion of feasibility of CT studies
at CEBAF.



\subsection{FSI effect dominates at
large transverse missing momenta}

Measuring the spectral function at large missing momenta $p_{m}$
is of great interest, because large $p_{m}$ are expected to give
a direct handle on the short-range correlations (SRC) of nucleons,
which is widely discussed for 30 years since seminal works by
Gottfried and Srivastava [43].
Our important finding [36]
is that in the transverse kinematics, the
FSI effect completely takes over the SRC effect.

The global effects of rescatterings are summarized in Table 1.
Here we present, for different values of $Q^{2}$, the fractions
$P^{(\nu)}=W^{(\nu)}/T_{A}$ of the $\nu$-fold rescatterings,
the nuclear transparency  $W^{(0)}$ for quasifree knockout in
parallel kinematics $p_{\perp}=0$, the total transparency $T_{A}$
and the average number of secondary rescatterings $\langle\nu\rangle$.
The $Q^{2}$ dependence of
$1-W^{(0)}$ is reminiscent of the energy dependence of
$\sigma_{tot}(pN)$, which is nearly flat at the kinetic energy
$T_{kin}\approx Q^{2}/2m_{p} \gsim 0.5$ GeV [42]. The $Q^{2}$
dependence of $1-T_{A}$ repeats the energy dependence of
$\sigma_{in}(pN)$, which rises rapidly up to
$T_{kin}\sim 1.5$ GeV and then approximately levels off [42].
The $Q^{2}$ dependence of the difference of the total and forward
transparency $T_{A}-W^{(0)}$ and of the multiplicity of secondary
rescatterings $\langle \nu \rangle$ repeats the energy dependence
of the elastic cross section
$\sigma_{el}(pN)=\sigma_{tot}(pN)-\sigma_{tot}(pN)$, which steadily
decreases with $T_{kin}$ [42].
The difference between $W^{(0)}$ and $T_{A}$ is a
convenient measure of the strength of rescatterings,
and in the CEBAF range
of $Q^{2}$ this difference is very large.

For our numerical estimates of the $p_{\perp}$-distribution
we use a simple, yet realistic,
parameterization of the SPMD [44]
\beq
n_{F}(\vec{k}) \propto
\exp\left(-{5\over 2}{k^{2}\over k_{F}^{2}}\right) +
\epsilon_{0}
\exp\left(-{5\over 6}{k^{2}\over k_{F}^{2}}\right) \, ,
\label{eq:4.2.1}
\endeq
where $\epsilon = 0.03$ and
the value of the Fermi-momentum $k_{F}$ has been taken from
Ref.~[39]: $k_{F}(C)=221\,{\rm MeV/c}$ and
$k_{F}(Pb)=265\,{\rm MeV/c}$.
 This parameterization is consistent with the
results from the y-scaling analysis [45]. The steeply
decreasing first term corresponds to the mean-field component of
the SPMD, while the second term describes the SRC tail.
In Fig.~17 we present our predictions for
$f_{A}(\vec{p}_{\perp})/
T_{A}$ for the $^{12}C(e,e'p)$ and $^{208}Pb(e,e'p)$ reactions at
$Q^2=1,2,4$ (GeV/c)$^2$, respectively.
The FSI contribution to $f_{A}(\vec{p}_{\perp})$
starts to dominate over the PWIA component $
P^{(0)}n_{F}(p_{\perp})$ already at
$p_{\perp}\gsim   350$ MeV/c for carbon
and  $p_{\perp}\gsim   300$ MeV/c for lead, which
is precisely the region thought of being dominated by
the SRC tail in the single-particle momentum distribution.

An obvious signal of rescattering is the multinucleon emission
(MNE). Rescatterings are not imperative for MNE, but
rescatterings leading to $p_{\perp} \gsim k_{F}$ are necessarily
followed by MNE. The contribution of rescatterings to
MNE is characterized by the average number of
secondary rescatterings $\langle \nu(p_{\perp})\rangle$,
which
is very large (Table 1). Evidently, the strength of MNE coming
from the rescattering mechanism must be the same in
the longitudinal and transverse cross sections, which is a
strong prediction. Presently, it can not
be tested for the lack of the experimental data
on large transverse missing momenta $p_{\perp} \gsim k_{F}$
taken at $Q^{2} \gsim $(1-2)GeV$^{2}$, which is the domain of
the forthcoming CEBAF experiments. This FSI driven MNE
contributes
significantly to the spectral function at large missing energy
$E_{m}$. Namely, the contribution $\Delta E_{m}$ to the missing
energy $E_{m}$ from the kinetic energy of
the recoil nucleons can
be estimated as
\beq
\Delta E_{m}\approx
 \left<{\vec{s}^{2}\over 2m_{p}}\right>=
 (1 + \epsilon_{0}) {k_{F}^{2}\over 5m_{p}} +
{p_{\perp}^{2}\over 2m_{p}}  \, .
\label{eq:4.2.2}
\endeq
Incidentally, the SRC mechanism leads to a
similar relationship between
$\Delta E_m$ and $p_{\perp}$. The rescattering effect
dominates the tail of the missing-energy distribution
\beq
{df_{A} \over dE_{m} } \propto
\exp\left(-{E_{m} \over E_{0}} \right)  \, ,
\label{eq:4.2.3}
\endeq
where the slope
\beq
E_{0} \approx {1\over 2Bm_{p}} \approx
{8\pi \sigma_{el}(pN)\over m_{p}\sigma_{tot}(pN)^{2}}\, .
\label{eq:4.2.4}
\endeq
The definitive signature of the FSI mechanism is a strong
$Q^2$ dependence of the slope $E_{0}$, which decreases from
$E_{0} \sim 250$ MeV at $Q^{2} \sim 1$ (GeV/c)$^{2}$ down to
$E_{0} \sim 90$ MeV at $Q^2 \sim 6$ (GeV/c)$^{2}$. Such a large
value of $E_{0}$ shows that in order to exhaust the closure, one
has to integrate up to very large $p_{\perp}$ and very large
missing energies $E_{m} \gsim E_{0}$.



\subsection{Rescattering effect in $d(e,e'p)$}

The rescattering effect is quite strong even in such a diluted
target as the deuteron.
The momentum distribution $f_d(\vec p)$ of the observed protons
equals [37]
\beq
f_d(p_{m,z},\vec{p}_{\perp})=
\left|\phi_{d}(p_{m,z},\vec{p}_{\perp})-
{\sigma_{tot}(pn) \over 16\pi^{2}}\int d^{2}\vec{k}\,
\phi_{d}(p_{m,z},\vec{p}_{\perp}-\vec{k})
\exp\left(-{1\over 2}B\vec{k}^{2}\right)
\right|^{2} \, .
\label{eq:4.3.1}
\endeq
Here $\phi_{d}(\vec{k})$ is the momentum-space wave function
of the deuteron, and gives the undistorted single-particle
momentum distribution, the second term describes the rescattering
effect. The measured transparency  $T_{d}$ depends on the
acceptance $p_{max}$:
\beq
T_{d}=
{\int^{p_{max}}d^{2}\vec{p}_{\perp}f_d(\vec{p}_{\perp}) \over
\int^{p_{max}}d^{2}\vec{p}_{\perp}\phi_d(\vec{p}_{\perp})^{2}}
\label{eq:4.3.2}
\endeq
The attenuation effect $T_{d} < 1$ comes from the interference
of the undistorted and rescattering terms in (\ref{eq:4.3.1}).
If $R_{d}^{2}p_{max}^{2} \gg 1$, but $Bp_{max}^{2} \ll 1$, which is
the case for the NE18, then
\beq
1-T_{d} \sim {\sigma_{tot}(pn) \over 2\pi R_{d}^{2}} \sim 0.07\,\, ,
\label{eq:4.3.3}
\endeq
which is about twice the Glauber shadowing effect in the
$\sigma_{tot}(Nd)$. Our prediction for $T_{d}$ is shown in Fig.~18
and the agreement with the NE18 data [4] is good.

At still larger values of $p_{\perp}$, the spectral function will
be entirely dominated by square of the rescattering term in
(\ref{eq:4.3.1}), cf. Eqs.~(\ref{eq:4.1.5},\ref{eq:4.1.6}),
\beq
f_{d}(p_{m,z},p_{\perp}) \sim
{\sigma_{el}(pn) \over 4\pi R_{d}^{2}} |\phi_{d}(p_{m,z})|^{2}
B\exp(-Bp_{\perp}^{2})
\label{eq:4.3.4}
\endeq
It exhibits features of the rescattering contribution to
the spectral function which are common to all the nuclei:
\begin{itemize}

\item
The $p_{\perp}$ distribution has the broad tail with the small
slope, which is the diffraction slope of the $pN$ elastic scattering.

\item
The longitudinal momentum distribution is much narrower and is
given by the PWIA single-particle momentum distribution.

\end{itemize}
In the opposite to that, the SRC effect in the single-particle
momentum distribution is an isotropic function
of the missing momentum $\vec{p}_{m}$, which must enable the
experimental separation of the isotropic SRC effect from the
sideways contribution from FSI.



\subsection{The theoretical interpretation of the NE18 data
$A(e,e'p)$ scattering}

The most important conclusion from the above discussion is that
the distortion effects make absolutely
illegitimate a factorization
of the measured spectral function $S(E_{m},\vec{p}_{m})$
into the PWIA spectral function $S_{PWIA}(E_{m},\vec{p}_{m})$
and an attenuation factor which is independent of the missing
energy and momentum.
In order to disentangle the genuine attenuation effect
from the distortion effects, one can define
nuclear transparency $T_{A}$ as the ratio of the experimentally
measured, and the theoretically calculated PWIA, cross sections
integrated over the experimental acceptance domain $D(NE18)$ in the
$(E_{m},p_{m,z},p_{\perp})$ space:
\beq
T_{A}(NE18)=
{\int_{D(NE18)} dE_{m}\, dp_{m,z}\,dp_{\perp}
S(E_{m},p_{m,z},p_{\perp}) \over
\int_{D(NE18)} dE_{m}\, dp_{m,z}\,dp_{\perp}
S_{PWIA}(E_{m},p_{m,z}+\Delta p_{m,z},p_{\perp})}
\label{eq:4.4.1}
\endeq
In order to eliminate the spurious effect in $T_{A}(NE18)$ due
to the FSI generated asymmetry of the nuclear spectral function,
we strongly advocate including the effect of the shift
(\ref{eq:4.1.4}) in the calculation of the PWIA cross section.
It was not included in the NE18 analysis. The effect of the
shift (\ref{eq:4.1.4}) on $T_{A}$ depends on the
$p_{m,z}$-acceptance. It vanishes for the wide $p_{m,z}$-acceptance.
For the narrow acceptance centered at $p_{m,z}=p^{*}$, the
longitudinal asymmetry effect can be evaluated as
\beq
{T_{A}(\Delta p_{m,z}) \over T_{A}(\Delta p_{m,z}=0)}
\approx 1 +
{5\over 2}\cdot {(\Delta p_{m,z}+p^{*})^{2}-(p^{*})^{2}
 \over k_{F}^{2} }
\label{eq:4.4.2}
\endeq
and can be quite strong at large $p^{*}$. At $p^{*}=0$, which is
approximetely the case in the NE18 experiment, the asymmetry effect
enhances $T_{A}$ by $\lsim 2\%$  and is still
within the error bars of the NE18 experiment.

The difference between $\sigma_{tot}(pN)$
and $\sigma_{in}(pN)$, and between $W^{(0)}$ and $T_{A}$ thereof,
is very large at moderate energies. $T_{A}(NE18)$ includes
partly the elastically rescattered struck protons. Since the
diffraction slope B rises with the proton energy,
the higher is $Q^{2}$ the larger is the fraction of the elastically
rescattered struck protons included in $T_{A}(NE18)$.
In Fig.~18
 we present our predictions [38] for $T_{A}$, $W^{(0)}$ and
$T_{A}(NE18)$ in which the $p_{\perp}$ integration for heavy nuclei is
extended up to $p_{max}=250\, {\rm MeV/c}$ as relevant to the NE18
situation [4]. We find very good quantitative agreement with the
NE18 data. The principle conclusion
from this comparison is that there is no large signal of CT in the
$A(e,e'p)$ scattering at $Q^{2} \leq 7$ GeV$^{2}$.

For the $^{12}C$ target we also show the effect of CT
which is still small even at the largest $Q^{2}$ of the
NE18 experiment (Fig.~18b), which must be contrasted with
large CT effect in the E665 experiment.
What makes the onset of CT in exclusive
 vector meson production and
$(e,e'p)$ scattering so much different?



\section{The onset of CT in $A(e,e'p)$ scattering}


\subsection{The ejectile state has a large size}

The $(e,e'p)$ scattering can be viewed as an absorption of
the virtual photon by the target proton, which leads to the
formation of the ejectile state $|E\rangle$, which then is
projected onto the observed final state proton. In the
$A(e,e'p)$ scattering, this ejectile state propagates in the
nuclear medium before evolving into the observed proton.
The strength of FSI depends on what is the transverse size
and the interaction cross section of the ejectile state.

In the exclusive production of vector mesons, the wave
function of the ejectile state equals [9,21]
\beq
\Psi_{E}(r) \propto \sigma(r)\Psi_{\gamma^{*}}(r)
\label{eq:5.1.1}
\endeq
Because of CT, in this case the ejectile wave function
$\Psi_{E}(r)$ has a hole at $r=0$, but at large $Q^{2}$ the
decrease of the wave function of the photon (\ref{eq:2.2.3})
takes over and $\Psi_{E}(r)$ has a small size
$\sim r_{S}$. Notice, that
 in terms of the ejectile state $|E\rangle$
the strength of FSI can be rewritten as
$\Sigma_{V}= \langle V|\sigma(r)|E\rangle/\langle V|E\rangle$.

The wave function of the ejectile state in the quasielastic
scattering of electrons is much simpler. It can be found in any
textbook in quantum mechanics and/or nuclear/particle physics.
Indeed, the charge form factor $G_{em}(Q)$
of the $q\bar{q}$ "proton" can be
written as (here the $\vec{r}$-plane is normal to the momentum
transfer $\vec{Q}$):
\beq
G_{em}(Q)=\langle p|E\rangle=
\int dz d^{2}\vec{r}\, \Psi_{p}^{*}(z,r)
\Psi_{E}(\vec{r},z)=
\int dz d^{2}\vec{r} \,\Psi_{p}^{*}(z,r)
\exp\left({i\over 2}Qz\right)\Psi_{p}(\vec{r},z)\, .
\label{eq:5.1.2}
\endeq
Therefore, the ejectile wave function
\beq
\Psi_{E}(\vec{r},z)=
\exp\left({i\over 2}Qz\right)\Psi_{p}(\vec{r},z)\, .
\label{eq:5.1.3}
\endeq
Because
$|\Psi_{E}(\vec{r},z)|^{2}= |\Psi_{p}(\vec{r},z)|^{2}$,
the ejectile
wave packet has the transverse size
{\bf identical} to the size of
the proton [46,13]. This result is readily generalized to
the three-quark proton and to the relativistic lightcone formalism.
The often made statement that the ejectile has a small size
  $\propto
1/\sqrt{Q^{2}}$ is quite misleading. The small size only is
believed to gradually appear when the ejectile wave function
is projected at very large $Q^{2}$ onto the observed proton [46,47],
but one must bear in mind the well known warning by
Isgur and Llewellyn Smith [48] that the onset of the small-size
dominance in the charge form factor is very slow. Even provided
that the proton form factor is dominated by the small size
contribution, it still
would be erroneous to identify the size of the
ejectile state with the region of $r$ which contribute to the
form factor of the proton. In the problem
of interest it is the large-size ejectile state with the initial
wave function (\ref{eq:5.1.3}) which starts propagating and
having FSI in the nuclear medium, evaluations of CT effects in
models which start with the ejectile of vanishing size must
be taken with grain at salt.



\subsection{Color transparency sum rules and vanishing FSI:
conspiracy of hard and soft scattering}

The ejectile state can be expanded in terms of the electroproduced
states $|i\rangle$ and the form factors $G_{ip}(Q)$ of the
$e\,p\rightarrow e\,i$ transitions:
\beq
|E\rangle = J_{em}(Q)|p\rangle=
\sum_{i}|i\rangle
\langle i|J_{em}(Q)|p\rangle =
\sum_{i}G_{ip}(Q)|i\rangle
\label{eq:5.2.1}
\endeq
where $G_{ip}(Q)$ are the $ep\rightarrow ei$ transition
form factors. Detailed description of how CT emerges in the
coupled-channel multiple scattering theory is given in [7,8,13,46].
There is much similarity to the treatment of CT in the production
of vector mesons. To the leading order in FSI,
\arr
T_{A} =
1-\Sigma_{ep}(Q){1\over 2A} \int d^{2}\vec{b}\,T(b)^{2}+... \, ,
\label{eq:5.2.2}
\endarr
where the strength of FSI is given by
\beq
\Sigma_{ep}(Q) =  {\langle p|\hat{\sigma}|E\rangle \over
\langle p|E\rangle}=
{1 \over G_{em}(Q)}\sum_{i}\sigma_{pi}G_{ip}(Q)=\sigma_{tot}(pp)+
{1 \over G_{em}(Q)}\sum_{i\neq p}\sigma_{pi}G_{ip}(Q)
\label{eq:5.2.3} \, .
\endeq
Here $G_{em}(Q)=G_{pp}(Q)$ is the form factor of the elastic
$e\,p$ scattering, $\hat{\sigma}$ is the cross
section or diffraction operator, which gives the forward
diffraction scattering amplitudes $M(jp\rightarrow kp)=
i\langle j|\hat{\sigma}|k\rangle = i\sigma_{jk}$.
The normalization is such that $\sigma_{tot}(jN)=\sigma_{jj}=
{\rm Im}M(jN\rightarrow jN)$.
Weak FSI and/or CT effect only is possible if the off-diagonal
contributions in Eq.~(\ref{eq:5.2.2}) cancel the $\sigma_{tot}(pN)$,
which comes from the conventional Glauber rescattering of the
struck proton. Evidently, such a cancellation {\it \bf requires a
very special conspiracy} between the electromagnetic form factors of
hard $e\,p$ interaction, and amplitudes of soft, forward,
diffractive hadronic transitions $i\,p \rightarrow p\,p$. Now we
shall demonstrate that such a conspiracy takes place.

In QCD the (anti)quarks in the hadron always have the one-gluon
exchange Coulomb interaction at short distances. This Coulomb
interaction leads to a very special asymptotics of the
form factor at $Q^{2} >> R_{p}^{2}$ [46,47]
(for the sake of simplicity
we discuss the two-quark state in the nonrelativistic
approximation)
\beq
G_{ik}(\vec{Q})
\propto {V(\vec{Q}) \over \vec{Q}^{2}}\Psi_{i}^{*}(0)\Psi_{k}(0)
\label{eq:5.2.4}
\endeq
Here $\Psi_{i}(0)$ are the wave functions at the origin, and
$V(Q)$ is the one-gluon exchange quark-quark hard
scattering amplitude.
Making use of the QCD asymptotics (\ref{eq:5.2.4}) we find [13]
\arr
\Sigma_{ep}(Q) = {1 \over G_{em}(Q)}
\sum_{i}\sigma_{pi}G_{ip}(Q)
\propto {1\over G_{em}(Q)}{V(\vec{Q}) \over \vec{Q}^{2}}
\sum_{i}\sigma_{pi}\Psi_{i}^{*}(0)\Psi_{p}(0)
\propto \sum_{i}\sigma_{pi}\Psi_{i}^{*}(0)   \, .
\label{eq:5.2.5}
\endarr
Remarkably, the {\sl r.h.s.} of Eq.~(\ref{eq:5.2.5})
vanishes by virtue of CT sum
rules [7]. The proof goes as follows:
The state of transverse size $\vec{r}$ can
be expanded in the hadronic basis as
$
\left.|\vec{r}\right>=
\sum_{i}\Psi_{i}(\vec{r})^{*}\left.|i\right> $
and the hadronic-basis expansion for $\sigma(r)$ reads  as
\beq
\sigma(\rho)=\langle \vec{r}|\hat{\sigma}|\vec{r}\rangle =
\sum_{i,k}
\left<\vec{r}|k\right>
\left<k|\hat{\sigma}|i\right>\left<i|\vec{r}\right>=
\sum_{i,k} \Psi^{*}_{i}(\vec{r})\Psi_{k}(\vec{r})
\sigma_{ki}
\label{eq:5.2.6}
\endeq
By virtue of CT $\sigma(r=0)=0$, and we obtain
the "CT sum rule" [7]
\beq
\sum_{i,k} \Psi_{k}(0)^{*}\Psi_{i}(0)\sigma_{ki} = 0      \, .
\label{eq:5.2.7}
\endeq
Considering the matrix elements $\langle \vec{r}|\hat{\sigma}|k\rangle$
at $\vec{r} \rightarrow 0$, one readily finds a whole family of CT
sum rules [7,13]
\beq
\sum_{i} \Psi_{i}(0)\sigma_{ik} = 0      \, .
\label{eq:5.2.8}
\endeq
It is precisely the sum rule (\ref{eq:5.2.8}), which in conjunction
with the QCD asymptotics of the electromagnetic form factors
(\ref{eq:5.2.4}) ensures that the strength of FSI (\ref{eq:5.2.3})
vanishes and CT indeed takes place.



\subsection{The onset of CT and the coherency constraint}

As we learned in section 3.4, at finite energy only few
intermediate states contribute coherently into $\Sigma_{V}$.
The case of the $\Sigma_{ep}$ is quite similar.
{}From the kinematics  of deep inelastic scattering
\beq
m_{i}^{2}-m_{p}^{2}=2m_{p}\nu - Q^{2}-2\nu k_{z}\, ,
\label{eq:5.3.1}
\endeq
where $k_{z}$ is the longitudinal momentum of the target nucleon.
Consequently, different components $|i\rangle$  of the ejectile
wave packet (\ref{eq:5.2.1}) are produced with the longitudinal
momenta differing by [7]
\beq
\kappa_{ip}=m_{p}{ m_{i}^{2} -m_{p}^{2} \over Q^{2}} =
{ m_{i}^{2} -m_{p}^{2} \over 2\nu }       \, ,
\label{eq:5.3.2}
\endeq
and acquire the relative phase $\kappa_{ip}(z_{2}-z_{1})$ during
propagation from the ($z_{1}$) production to the ($z_{2}$)
rescattering point. Then, repeating considerations which lead to
Eqs.~(\ref{eq:3.4.10}, \ref{eq:3.4.11}), we find
[7,8,13]
\arr
\langle p|\hat{\sigma}J_{em}|p\rangle =
\sum_{i}\langle p|\hat{\sigma}|i\rangle\langle i|J_{em}|p\rangle
{}~~~~~~~~~~~~~~~~~~~~~~~~~~~
\nonumber\\
\Longrightarrow
\sum_{i}\langle p|\hat{\sigma}|i\rangle\langle i|J_{em}|p\rangle
\exp[i\kappa_{ip}(z_{2}-z_{1})]
\Longrightarrow
\sum_{i}\langle p|\hat{\sigma}|i\rangle\langle i|J_{em}|p\rangle
G_{A}^{2}(\kappa_{ip})
\label{eq:5.3.3}
\endarr
and
\beq
\Sigma_{ep}(Q)=\sigma_{tot}(pN)+
\sum_{i\neq p}{G_{ip}(Q) \over G_{em}(Q)}
\sigma_{ip}G_{A}(\kappa_{ip})^{2}
\label{eq:5.3.4}
\endeq
At small $Q^{2}$ such that
\beq
Q^{2} \ll   2R_{A}m_{p}\Delta m \sim
(3-5)\cdot A^{1/3} {\rm GeV}^{2}\, ,
\label{eq:5.3.5}
\endeq
for all inelastic channels $G_{A}(\kappa_{ip}) \ll 1$, so that
$\Sigma_{ep}=\sigma_{tot}(pN)$, and FSI measures the free-nucleon
cross section. With increasing $Q^{2}$, for larger and larger
number of inelastic intermediate states channels $G_{A}(\kappa_{ip})
\approx 1$, the CT sum rule (\ref{eq:5.2.8}) will be better
saturated and $\Sigma_{ep}(Q) \rightarrow 0$.



\subsection{What makes the onset of CT in $A(e,e'p)$ and vector
meson production different?}

In the vector meson production, it is the energy $\nu$ which
controls the number of intermediate states which contribute
coherently to $\Sigma_{V}$ and the signal of CT. The virtuality
$Q^{2}$ is an independent parameter, which controls the srength
of destructive interference between the diagonal (elastic) and
off-diagonal (inelastic) contributions to $\Sigma_{V}$. In the
quasielastic scattering of electrons $Q^{2} \approx 2m_{p}\nu$,
and one can increase the energy $\nu$ only at the expense of
very high $Q^{2}$.

There is one more fundamental difference between the two
processes. Compare in more detail the $\Sigma_{J/\Psi}$
Eq.~(\ref{eq:3.4.10}) and $\Sigma_{ep}(Q)$ Eq.~(\ref{eq:5.3.4}).
Notice, that $M(J/\Psi\,N\rightarrow \Psi'N) =
\sigma_{J/\Psi\,\Psi'}$ and, in the generic case, the off-diagonal
matrix elements $\sigma_{ip}$ are typically smaller than the
diagonal $\sigma_{ii}=\sigma_{tot}(i\,N)$. The difference comes
from the two ratios
\beq
{M(\gamma^{*}N \rightarrow \Psi'N)\over
M(\gamma^{*}N\rightarrow \Psi'N)}~~~~~~{\rm and}~~~~~~
{G_{ip}(Q)\over G_{em}(Q)}  \, ,
\label{eq:5.4.1}
\endeq
which play the identical role in Eqs.~(\ref{eq:3.4.10}),
(\ref{eq:5.3.4}), respectively. In the vector meson production,
the ratio (\ref{eq:5.4.1}) increases with $Q^{2}$, which
further enhances the off-diagonal contribution to $\Sigma_{J/\Psi}$
and the signal of CT thereof. In the case of $A(e,e'p)$ at
large $Q^{2}$, the ratio $G_{ip}(Q)/G_{em}(Q)$ does not change
with $Q^{2}$, see Eq.~(\ref{eq:5.2.4}) and Stoler's review [49].



\subsection{The realistic estimates of the signal of CT
in $A(e,e'p)$ scattering}

One needs to evaluate the off-diagonal contributions to
$\Sigma_{ep}$. Let us accept the optimstic scenario that
the form factor ratios $G_{ip}(Q)/G_{em}(Q)$ are precociously
close to QCD asymptotic predictions (\ref{eq:5.2.4}), and
concentrate on the off-diagonal diffraction production
amplitudes $\sigma_{ip}$. The experimentally observed mass
spectrum of the diffraction excitation of protons is shown
in Fig.~19 (for the review see [50]). Because of the
nuclear form factor $G_{A}(\kappa_{ip})$, the onset of CT
signal starts with the contribution to $\Sigma_{ep}$ from
the low-mass states.

In the diffractive mass spectrum of Fig.~19 there is a
prominent signal of excitation $pp\rightarrow (\pi N)p$ of
the low-mass $\pi N$ state, which dominates at
$M^{2} \lsim 2$GeV$^{2}$. Its origin is in the admixture
of the $\pi (3q)$ Fock state in the lightcone proton [13,51]
\beq
 |N\rangle = \cos\theta|3q\rangle + \sin \theta|3q +\pi\rangle \,  ,
\label{eq:5.5.1}
\endeq
where the mixing angle $\theta$ is related to the number of
pions $n_{\pi}$ in the nucleon as
\beq
\sin^{2}\theta= {n_{\pi}\over 1+n_{\pi}}
\label{eq:5.5.2}
\endeq
There is a mounting evidence for such a hybrid quark core-pion
structure of the nucleon, coming from the violation of the
Gottfried Sum Rule and the recent NA51 observation [52] of large
$\bar{u}/\bar{d}$ asymmetry in the proton sea in the Drell-Yan
production [53]. The diffractive excitation of the $N+\pi$ Fock
component of the nucleon (known since long also as the
Drell-Hiida-Deck process [50]), gives an excellent description
of the $\pi N$ mass spectrum. For the purposes of the present
analysis it is important, that the $\pi N$ state contributes to
$\Sigma_{ep}$ a term [13]
\beq
\Delta\Sigma_{ep}(\pi N) \approx -\sigma_{tot}(\pi N)
{n_{\pi} \over 1+n_{\pi}} \approx -8{\rm mb}
\label{eq:5.5.3}
\endeq
This result can be interpreted as follows. The pion-baryon component
of the nucleon has a large size and a rapidly decreasing form factor,
so that only the baryonic $3q$ core of the nucleon contributes to
the ejectile state (the detailed analysis of electromagnetic form
factor of the hybrid proton is given in [51]). The interaction
cross section of the $3q$ core stripped off of pions, is smaller
that $\sigma_{tot}(pN)$ by precisely the amount (\ref{eq:5.5.3}).
Notice, how the nonperturbative quark core-pion Fock component of
the nucleon gives a unified description of
\begin{itemize}
\item
the diffraction
dissociation of nucleons in the soft diffractive scattering,
\item
the asymmetry of the quark-antiquark sea in protons as probed
in deep inelastic scattering,
\item
the onset of color transparency in quasielastic scattering of
electrons on nuclei.
\end{itemize}

Diffraction excitation of higher mass states comes from the
excitation of the $3q$ core of the nucleon. In [7] we described
interaction of the $3q$ core in a QCD approach based on the dipole       cross
section [15], qualitatively shown in Fig.~1. The resulting
diffraction matrix $\sigma_{ik}$ has the built-in CT property and
satisfies CT sum rules. The proton and its excitations are
described by the diquark-quark harmonic oscillator model.
Within the model, diffraction dissociation is dominated by
excitation of $N^{*}$ states of the positive parity shells.
We find a good parameter-free description [54] of the diffractive
mass spectrum shown in Fig.~19,
and rightfully expect the realistic
estimate of CT effects in $\Sigma_{ep}(Q^{2})$. The excess of
the observed mass spectrum in the so-called triple-pomeron
mass region $M^{2} \gsim 10$GeV$^{2}$ is an effect of excitation
of higher $qqqg_{1}..g_{n}$ Fock states [17], which was not
included in [7]. However, in the practically accessible range
of $Q^{2}$, CT effect is completely dominated by excitation of
the $\pi N$ states and resonances of the first excited
positive-parity shell $L=1$, with little contribution of states
of higher shells $L \geq 2$.

The resulting predictions [54] for $\Sigma_{ep}(Q)$ are shown
in Fig.~20. With increasing $Q^{2}$, the strength of FSI
decreases much slower than $\propto 1/Q^{2}$: the increase
of $Q^{2}$ by one order in magnitude from 5 to 50 GeV$^{2}$
is followed by decrease of $\Sigma_{ep}(Q)$ only by the
factor 3 from 36mb to 12 mb, respectively.
The CT effect in $T_{A}(Q^{2})$ follows closely
the variation of $\Sigma_{ep}(Q^{2})$. The heavier is the
nucleus, the larger $Q^{2}$ is needed for the onset of CT.
The values of
$Q^{2}$ studied by NE18 correspond to the threshold of CT for
the carbon nucleus, already at  $Q^{2} \sim 20$ GeV$^{2}$ we predict
a substantial CT effect.



\section{CT in $A(e,e'p)$ scattering at CEBAF}



\subsection{$^{4}He(e,e'p)$ scattering:  CEBAF's choice }

The principal inhibitor of the precocious CT is the coherency
constraint, quantified by the factor $G_{A}(\kappa_{ip})^{2}$
in the off-diagonal contribution to $\Sigma_{ep}(Q)$. In order
to have large CT effect one must excite intermediate
states of high mass $m_{i}$. The condition $\kappa_{ip}R_{A}
\lsim 1$ implies, that excitation of the same mass on different
nuclei requires
\beq
Q^{2} \gsim (m_{i}^{2}-m_{p}^{2})R_{A}m_{p}
\label{eq:6.1.1}
\endeq
Consequently, one can
hope to relax the coherency constraint and enhance the CT effect
going to lighter nuclei which have smaller radius and less
steeply decreasing charge form factor $G_{A}(\kappa)$.

With allowance for the center-of-mass motion, for the light
nuclei [13,54]
\beq
\Sigma_{ep}(Q^{2})=\sigma_{tot}(pN)+\sum_{i\neq p}\sigma_{ip}
{G_{ip}(Q^{2})\over G_{pp}(Q^{2})}
G_{A}^{2}({A\over A-1}\kappa_{ip}^{2})
\, .
\label{eq:6.1.2}
\endeq
The factor $A/(A-1)$ in the nuclear form factor, which
comes from the center-of-mass motion important in light
nuclei, is the most unwelcome news. For the deuteron
target, it makes the
onset of CT slower than for the $^{12}C$ target. The $^{3}He$
nucleus has a size smaller than the $^{12}C$ nucleus, but the
onset of CT in both cases will be equally slow, which makes
the $^{3}He(e,e'p)$ reaction
ill-suited for the search of CT effects at
CEBAF. Only for the $^{4}He$ nucleus, the onset of CT will be
sooner than for the $^{12}C$ nucleus, which makes $^{4}He$ the
CEBAF's choice target.

In the $^{4}He(e,e'p)$ scattering, nuclear attenuation effect
is typically $1-T_{A} \sim 0.25$ [55] and with the onset of
CT it will decrease
\beq
1-T_{A} \propto
\eta = {\Sigma_{ep}(Q)\over \sigma_{tot}(pN)}\,.
\label{eq:6.1.3}
\endeq
Our predictions [54] of $\eta$ for $^{4}He(e,e'p)$
scattering are shown in Fig.~21. We also show the decomposition
of the CT signal $1-\eta$ in terms of excitation of
inelastic intermediate states. At $Q^{2} \lsim $(5-6)GeV$^{2}$
the signal of CT comes primarily from the $\pi N$
intermediate states. In this case the reduction of
$\Sigma_{ep}(Q)$ from $\sigma_{tot}(p N)=$40mb measures how
much the size of the quark core of the nucleon stripped off the
pionic cloud, is smaller than the size of the physical proton.
With the increasing $Q^{2}$, excitation of the intermediate
nucleonic resonances starts contributing to the signal of CT
and atkes over at $Q^{2} \gsim $7GeV$^{2}$.
The experimental observation of the CT predicted $\sim 20\%$
decrease of $\eta$ at $Q^{2}=6$GeV$^{2}$, requires the
measurements of $T_{A}$ with $\lsim 1\%$ statistical accuracy.
This challenge can be met at CEBAF thanks to its high luminocity.



\subsection{The signal of CT at large transverse missing
momenta}

The broadening of the $p_{\perp}$ distribution discussed in
sections 4.1-4.3 is generated by FSI and must vanish in the
CT regime.
With allowance for inelastic intermediate states,
the propability of $\nu$-fold rescattering $W^{(\nu)}$ will
still be given by Eq.~(\ref{eq:4.1.7}) subject to the
substitution [13,14]
\beq
\sigma_{el}(pN)^{\nu} \Longrightarrow
\sigma_{el}(pN)^{\nu}
\left[{\langle p|\hat{\sigma}^{\nu} |E\rangle
\over \langle p|E\rangle
\sigma_{tot}(pN)^{\nu} }\right ]^{2} \, .
\label{eq:6.2.1}
\endeq
In the case of $^{4}He(e,e'p)$ scattering, the tail of the
$p_{\perp}$-distribution $f_{A}(p_{\perp})$ will be dominated
by single scattering term, which will be proportional to the
factor
\beq
W_{1} \propto  \eta^{2}=
\left({\Sigma_{ep}(Q^{2})\over \sigma_{tot}(pN)}\right)^{2}
\label{eq:6.2.2}
\endeq
Our prediction [54]
for the rescattering
suppression factor $\eta^{2}$ is shown
in Fig.~21. Here the CT effect is twice as large compared to
nuclear transparency $T_{A}$. One has to pay a heavy price,
though: very accurate theoretical calculation of the
rescattering effect in the absence of CT is a must before one
can claim the reliable experimental determination of
this suppression factor and separation of the CT effect.
Such a calculation must be very accurate, indeed, because
the magnitude and the $p_{\perp}$-dependence of the differential
cross section of elastic $pN$ scattering changes very rapidly over
the range of $Q^{2}$ of the interest.

Furthermore, the rescattering component of $f_{A}(p_{\perp})$
appears on the background from the short-range correlation
effect, see Fig.~17,  and very good understanding of this
background is necessary for reliable determination of a small
decrease of $\eta^{2}$ with the increase of $Q^{2}$.
The coincidence detection of the recoil nucleon little helps
in this respect: once the struck proton emerges with large
$p_{\perp} \gg k_{F}$, it will always recoil against the
second nucleon emitted with the momentum
$\vec{p}_{2} \approx -\vec{p}_{m}$, and one can not tell
whether the large $p_{\perp}$ originated from the correlation
of two nucleons at short distances, or from the rescattering
process. For this reason, the theoretical interpretation of
high-precision measurements of the $Q^{2}$ dependence of
nuclear transparency $T_{A}$ will be much easier, compared to
the $Q^{2}$ dependence of the large-$p_{\perp}$ production
and/or the $^{4}He(e,e'pp)$ reaction.



\subsection{Fermi-motion and asymmetry effect in nuclear
transparency}

In the $(e,e'p)$ scattering, the longitudinal momentum $k_{z}$
of the target nucleon is uniquely determined by the electron
scattering kinematics:
\beq
x={Q^{2}\over 2m_{p}\nu} =1+{k_{z}\over m_{p}}\, .
\label{eq:6.3.1}
\endeq
By kinematics of the electroproduction  Eq.~(\ref{eq:5.3.1}),
the different components of the ejectile wave packet are produced
on the target nucleon having different longitudinal Fermi
motion, see Eq.~(\ref{eq:5.3.2}). Consequently, varying the
Bjorken variable $x$, one can change the composition of the
ejectile wave packet [8,56,57], which gives a new handle on the CT
effect. A detailed theory of the effect is given in [8].

Following West,
it is convenient to introduce the PWIA structure function
\beq
F_{A}(x)=\int d^{3}\vec{k}{dn\over d^{3}\vec{k}} \,
\delta(1+{k_{z}\over m_{p}}- x) =
{1\over 2\pi}
\int dk_{z} \,
\delta(1+{k_{z}\over m_{p}}- x)
\int dz \,
\rho(0,z)
\exp(ik_{z}z)  \, .
\label{eq:6.3.2}
\endeq
Then, the Fermi-motion effect is quantified by the $x$-dependent
decomposition of the ejectile state
\beq
|E\rangle_{eff} =
\sum_{i}
G_{ip}(Q)|i\rangle
{F_{A}(x+{1\over 2}\Delta x_{ip})\over F_{A}(x)}
G_{A}(\kappa_{pi})
\label{eq:6.3.3}
\endeq
and the $x$-dependent strength of FSI
\arr
\Sigma_{ep}(x,Q^{2})=
\sigma_{tot}(pN)+
\sum_{i\neq p} \sigma_{pi}
{G_{ip}(Q)\over G_{em}(Q)}
{F_{A}(x+{1\over 2}\Delta x_{ip})\over F_{A}(x)}
G_{A}(\kappa_{ip})^{2}   \, ,
\label{eq:6.3.4}
\endarr
where
\beq
\Delta x_{ip}={\kappa_{ip}\over m_{p}}=
{m_{i}^{2}-m_{p}^{2} \over Q^{2}}  \,\, .
\label{eq:6.3.5}
\endeq
In their counterpart of (\ref{eq:6.3.3}), instead of
$F_{A}(x_{Bj}+{1\over 2}\Delta x_{ip})/F_{A}(x)$
Jennings and Kopeliovich [56]
give $\sqrt{F_{A}(x+\Delta x_{ip})/F_{A}(x)}$, which is incorrect.

The $x$-dependence of $\Sigma(x,Q^{2})$ and of the transparency
$T_{A}(x,Q^{2})$  comes from the fact that
the ratio $F_{A}(x+{1\over 2}\Delta x_{ip} )/F_{A}(x)$ is
asymmetric  around $x=1$ [8,56,57].
 It is smaller than unity at
$x> 1$, which suppresses the inelastic contribution to
$\Sigma_{ep}$ and hence reduces the color transparency signal.
For $x < 1$, on the other hand, transparency is enhanced.
With $F_{A}(x)$ evaluated using the single-particle momentum
distribution (\ref{eq:4.2.1}), we find the results shown in
Fig.~22.
At small $Q^{2}$ the Fermi motion effect is
numerically small because the contributions from inelastic channels
to $\Sigma_{ep}(x,Q^{2})$ are supressed by
$G_{A}(\kappa_{ip})^{2}$. At higher $Q^{2}$ these contributions
increase and the $x$-dependence of
$T_{A}(x,Q^{2})$ becomes significant.
The SRC component of $F_{A}(x)$
has a significant influence on the Fermi motion effect
reducing the departure of the
ratio $F_{A}(x+{1\over 2}\Delta x_{ip} )/F_{A}(x)$ from unity in the
region of $x$ where $F_{A}(x)$ is dominated by this component.
This is demonstrated by the
dashed curves in Fig.~22, where $T_{A}(x,Q^{2})$ is evaluated
with $F_{A}(x)$ which includes only the mean-field component.
$T_{A}(x,Q^{2})$ has a peak and the signal of CT is
maximized at $x \sim 0.8$.

We emphasize that the asymmetry of nuclear transparency
about $x=1$ is the CT effect. It is large and persists at
large $Q^{2}$. Zooming at $x\sim 0.8$ will significantly
enhance the chances of discovering CT at CEBAF. However,
there is a strong background to this CT induced asymmetry,
which comes from the effective shift of the missing
momentum (\ref{eq:4.1.4}), which also generates the
asymmetry in nuclear transparency. Namely, $x\sim 0.8$
corresponds to $p^{*} \sim k_{F}$, and the FSI generated
asymmetry (\ref{eq:4.4.2}) is as large as the CT effect
shown in Fig.~22.


\section{Conclusions}

The E665 observation [5] of strong signal of CT in exclusive
production of vector mesons is a major breakthrough in the
subject of CT and paves the way for dedicated experiments
on CT. The E665 effect was predicted [9] (not postdicted!), and
the agreement betwen the theory and experiment is very good. The
nonobservation of CT in the NE18 experiment [4] was also
predicted [7,8] and is
a good confirmation of theory. We have a good understanding
of why the onset of CT in $A(e,e'p)$ and $\gamma^{*}A\rightarrow
VA,\,VA^{*}$ reactions is so strikingly different.

The purpose of this presentation was to make the strong case
for a broad experimental program on electroproduction of
light vector mesons $\rho^{0},\,\omega^{0},\,\phi^{0},
\rho',\,\omega',\,\phi'$ at CEBAF. Here the principal benefit
is the theoretically well understood shrinkage of the
virtual photon which elastically produces the vector meson.
Variations of the scanning radius are very strong already
at $Q^{2} \lsim$(1-3)GeV$^{2}$, and the first-class CT
experiments with large expected signal of CT can be performed
at the energy-upgraded CEBAF [58]. CT in conjunction with the
node structure of wave functions of radially excited vector
mesons $V'=\rho',\,\omega',\,\phi'$ leads to a very rich pattern
of anomalous $A$ and $Q^{2}$ dependence of the $V'$ production.
This unique possibility of directly scanning the wave function
of radial excitations at CEBAF must not be overlooked, and
CEBAF is the unique facility which can do the job. Furthermore,
the CEBAF experiments on radially excited vector mesons can
contribute much to the spectroscopy of vector mesons, which
presently leaves much to be desired [59].
Besides the above CT aspects,
there is much interest in studies of nuclear modifications
of properties of vector mesons in electroproduction on nuclei
at CEBAF [60].

$^{4}He$ emerges as CEBAF's choice target to
search for CT in $A(e,e'p)$ scattering.
The realistic, and conservative,
estimates show that CT effect in the conventional transmission
experiment can be observed at CEBAF. CT effect can be
enhanced studying the asymmetry of nuclear transparency (the
Fermi-motion effect) in the parallel kinematics. Here more
theoretical work is needed to reliably calculate the
background asymmetry
originating from the more conventional final state interaction.
CT effect aslo can be enhanced looking at large missing
momentum in the transverse kinematics. Here a very accurate
calculation of the rescattering effect and of the background
short-range correltion effect
is needed for a realiable
normalization of the cross section and extraction of CT effect.
\medskip\\
{\bf Acknowledgements}
\medskip\\
NNN thanks N.Isgur, R.Milner and P.Stoler for invitation
to this stimulating Workshop.
\pagebreak\\

\pagebreak


\setlength{\parindent}{0.0cm}
\begin{center}
{\bf Tables}
\end{center}

\begin{center}

Table 1.  Probabilities $P^{(\nu)}$
for $\nu$-fold rescatterings
(in per cent) for the $(e,e'p)$ reaction on $^{12}$C (upper block) and
$^{208}$Pb  (lower block). In addition we show the
nuclear transparency $W^{(0)}$ for quasifree knockout in
the parallel kinematics, the nuclear transparency for
the semi-exclusive reaction rate $T_{A}$ and  the
average number of rescattering $\langle \nu \rangle$.
\\
\end{center}
\vskip 2cm

\begin{center}

\begin{tabular}{|l|c|c|c|c|c|c|c|c|}
 \hline
 $ Q^2  $ &   0   &   1   &   2   &   3   &   4   & $W^{(0)}$ & $T_A$ &
 $\langle \nu \rangle$ \\
 \hline
 $1$   & 65.15 & 23.40 &  8.09 &  2.49 &  0.67 &  0.63 &  0.97 &  0.51 \\
$1.2$  & 68.37 & 22.06 &  6.96 &  1.98 &  0.50 &  0.60 &  0.87 &  0.45 \\
$1.5$  & 71.94 & 20.39 &  5.76 &  1.48 &  0.34 &  0.56 &  0.77 &  0.38 \\
 $2$   & 74.18 & 19.26 &  5.03 &  1.21 &  0.26 &  0.54 &  0.72 &  0.34 \\
 $4$   & 81.31 & 15.24 &  2.87 &  0.49 &  0.08 &  0.54 &  0.67 &  0.23 \\
 $6$   & 83.40 & 13.89 &  2.31 &  0.35 &  0.05 &  0.55 &  0.66 &  0.19 \\
 $8$   & 85.65 & 12.33 &  1.76 &  0.23 &  0.03 &  0.56 &  0.65 &  0.17 \\
       &       &       &       &       &       &       &       &       \\
 \hline
 $1$   & 30.13 & 21.60 & 16.09 & 11.69 &  8.12 &  0.25 &   0.84 &  1.98 \\
$1.2$  & 39.32 & 23.70 & 15.00 &  9.39 &  5.70 &  0.23 &   0.57 &  1.46 \\
$1.5$  & 49.36 & 24.44 & 12.81 &  6.72 &  3.46 &  0.20 &   0.41 &  1.03 \\
 $2$   & 55.08 & 24.12 & 11.22 &  5.25 &  2.43 &  0.19 &   0.34 &  0.83 \\
 $4$   & 67.45 & 21.54 &  7.30 &  2.48 &  0.83 &  0.19 &   0.28 &  0.49 \\
 $6$   & 70.66 & 20.39 &  6.23 &  1.91 &  0.57 &  0.20 &   0.28 &  0.42 \\
 $8$   & 74.37 & 18.79 &  5.02 &  1.34 &  0.35 &  0.20 &   0.27 &  0.35 \\
       &       &       &       &       &       &       &        &       \\
 \hline
\end{tabular}

\end{center}

\label{tab:contri}

\pagebreak

{\bf \Large Figure captions:}

\begin{itemize}

\item[Fig.~1 ]
- Qualitative dependence of the universal dipole cross
section  $\sigma(r)$ on  the dipole
size $r$ for scattering of the color dipole on the nucleon
target. Different processes probe this universal dipole
cross section at different $r$ [12].

\item[Fig.~2 ] -
The qualitative pattern of $Q^{2}$-dependent scanning
of the wave functions of the ground state $V$
and the radial excitation
$V'$ of vector meson [9,27]. The scanning distributions
$\sigma(r)\Psi_{\gamma^{*}}(r)$ shown by the solid and dashed
curve have the scanning radii $\rho_{S}$ differing by a factor 3.
All wave functions are in arbitrary units.

\item[Fig.~3 ] - Predictions [23] for the exclusive production
of vector mesons $\rho^{0},\phi^{0}$ vs. the NMC data [25].

\item[Fig.~4 ]
- Predictions [10] of
nuclear transparency $T_{A}=\sigma_{A}/A\sigma_{N}$
for the incoherent
exclusive production of $\rho^{0}$ mesons vs. the E665 data [5].

\item[Fig.~5 ] -
Predictions [10] of nuclear transparency
$T_{A}^{(coh)}/T_{C}^{(coh)}
=[144d\sigma_{A}/A^{2}d\sigma_{C}]_{\vec{q}\,^{2}=0}$
for the forward coherent production of the $\rho^{0}$ mesons.

\item[Fig.~6 ] -
Predictions [10] of the $Q^{2}$ dependence of the
ratio of cross sections
$R_{coh}(A/C)=
12\sigma_{A}/A\sigma_{C}$
for coherent production of the $\rho^{0}$ meson vs. the E665
data [5]. The arrows indicate predictions for the
complete CT (vanishing FSI).

\item[Fig.~7 ] -
Predictions [10] of the $Q^{2}$ dependence of exponents of
parametrizations $\sigma_{A}(inc) \propto A^{\alpha_{inc}}$,~~
$[d\sigma_{A}(coh)/dq^{2}]_{\vec{q}\,^{2}=0} \propto
A^{\alpha_{coh}(\vec{q}\,^{2}=0)}$ and $\sigma_{A}(coh) \propto
A^{\alpha_{coh}}$
vs. the E665
data [5].

\item[Fig.~8 ]
- Energy dependence of the ratio of transparency in quasielastic
$J/\Psi$ photoproduction on tin and carbon predicted [22] from
Eq.~(\ref{eq:3.1.7}) (shown by solid line) versus the NMC data
[28] for the 200 and 280 GeV primary muons [34]. The dotted
line shows the prediction of the classical expansion  model
by Farrar et al. [31].

\item[Fig.~9.] -
The predicted $Q^{2}$ and $\nu-$dependence of the nuclear transparency
in the virtual photoproduction of the $\rho^{0}$-mesons [9].

\item[Fig.~10] -
The predicted $Q^{2}$ and $\nu$-dependence of the nuclear
transparency in the virtual phtotoproduction of
the heavy quarkonia [9].
The qualitative pattern is the same from the light to heavy nuclei.
\item[Fig.~11 ] -
The $Q^{2}$ dependence of the $\rho'(2S)/\rho^{0}(1S)$ ratio of
forward production cross sections, which exhibits a dip because
of the exact node effect in the $\rho'(2S)$ production.

\item[Fig.~12 ] -
The $Q^{2}$ and $A$ dependence of nuclear transparency for the
$\rho^{0}(1S)$ and $\rho'(2S)$ electroproduction on nuclei.

\item[Fig.~13 ] -  Predictions [24] for the
onset of antishadowing in the incoherent
real photoproduction of radial excitation $\phi'(2S)$ opn
nuclear targets.

\item[Fig.~14 ] -  Predictions [24] for nuclear transparency
$T_{A}=\sigma_{A}/A\sigma_{N}$ in electroproduction of
the $\phi^{0}(1S)$ and $\phi'(2S)$ on nuclei.

\item[Fig.~15 ] -  Predictions [24] for the CT and
node-effect driven
steep $Q^{2}$ dependence of nuclear transparency in the
electroproduction of the $\phi'(2S)$ on nuclei.

\item[Fig.~16 ] -
The final-state interaction in quasielastic $(e,e'p)$
scattering. FSI regenerates the observed proton $|p\rangle$
from  electroproduced intermediate states $|i\rangle$.

\item[Fig.~17 ] - Predictions [36] for
the transverse missing
momentum distribution of
struck protons from  $^{12}C(e,e'p)$ (left hand side)
and $^{208}Pb(e,e'p)$ reactions (right hand side) for
different values of $Q^2$.
The solid lines correspond to the full momentum distributions.
The PWIA component of the missing momentum distribution
($\nu=0$) is shown by the long-dashed line.
The $\nu$-fold rescattering contributions are represented by
the short-dashed lines ($\nu=1$),
the dash-dotted lines ($\nu=2$) and the dotted lines ($\nu=3$)
respectively.

\item[Fig.~18 ] -
   Predictions [37] for the $Q^{2}$ dependence of nuclear
   transparency are compared with the NE18 data [4].
   The dashed curve denotes the $p_{\perp}$-integrated nuclear
   transparency $T_{A}$; the dotted curve gives the
   nuclear transparency
   $W^{(0)}$ measured in parallel kinematics at $p_{\perp}=0$;
   the solid curve represents the nuclear transparency $T_{A}(NE18)$
   including the NE18 acceptance cuts [4]; the
   dot-dashed curve in panel b)  shows the predicted [37]
   effect of the onset of CT.

\item[Fig.~19 ] - The predicted [54] mass spectrum (the solid
line) in the diffraction dissociation
of protons $pp\rightarrow pX$ into state of invariant mass $M$.
The data are taken from the compilation [50]. Contribution to
the mass spectrum from excitation of the $\pi N$ continuum is
shown by the dotted line; contributions from
excitation of resonances corresponding
to the positive parity excited shells is shown by dashed lines.

\item[Fig.~20 ] -
Predictions [54] for the $Q^{2}$-dependence of the strength of
FSI $\Sigma_{ep}(Q^{2})$ in $^{12}C(e,e'p)$ reaction (left box)
and of nuclear transparency $T_{A}$ for different nuclear targets.


\item[Fig.~21 ] - The top box:
Predictions [54] for the $Q^{2}$ dependence of
the scaled strength of FSI $\eta=\Sigma_{ep}(Q^{2})/\sigma_{tot}(pN)$
in $^{4}He(e,e'p)$ reaction (the solid line). Shown separately are
the contributions to $1-\eta$ which is the signal of CT, from
excitation of intermediate $\pi N$ state (the dotted line) and
from excitations of nucleonic resonances (the dashed line). \\
The bottom box:
   Predictions [54] for the CT effect in the rate of
   $^{4}He(e,e'p)$ reaction with high transverse missing
   momentum. The quantity shown is the relative
   strength of the elastic-rescattering tail in the
   $p_{\perp}$-distribution which is given by $\eta^{2}=
   [\Sigma_{ep}(Q^{2})/\sigma_{tot}(pN)]^{2}$

\item[Fig.~22 ] - The predicted [8]
dependence of nuclear transparency $T_{A}$ (upper part)
and of the observable $\Sigma_{ep}(x,Q^{2})$ (lower part)
on the Bjorken variable $x$ for different values of $Q^{2}$
(in $(GeV/c)^{2}$) listed in the figure. The dashed curves
are obtained if only the mean-field component of the
single-particle momentum distribution (\ref{eq:4.2.1})
is retained in evaluation of $F_{A}(x)$ in Eq.~(\ref{eq:6.3.2}).
top results
are for $^{12}C(e,e'p)$ scattering.


\end{itemize}
\end{document}